\documentclass[fleqn,usenatbib]{mnras}

\usepackage[T1]{fontenc}

\DeclareRobustCommand{\VAN}[3]{#2}
\let\VANthebibliography\thebibliography
\def\thebibliography{\DeclareRobustCommand{\VAN}[3]{##3}\VANthebibliography}

\usepackage{graphicx}	
\usepackage{amsmath}	

\title[Resolved $^{12}$CO(2-1) emission in ODISEA sources]{Gas distribution in ODISEA sources from ALMA long-baseline observations in $^{12}$CO(2-1)}

\author[J. Antilen et al.]{
Juanita Antilen,$^{1,2}$\thanks{E-mail: juani.antilenromero@gmail.com}
Simon Casassus,$^{1,2,3,4}$
Lucas A. Cieza,$^{2,5}$
Camilo González-Ruilova$^{2,5,6}$
\\
$^{1}$Departamento de Astronom\'{\i}a, Universidad de Chile, Casilla 36-D, Santiago, Chile\\
$^{2}$Millennium Nucleus on Young Exoplanets and their Moons (YEMS), Chile\\
$^{3}$Facultad de Ingenier\'ia y Ciencias, Universidad Adolfo Ib\'a\~nez, Av. Diagonal las Torres 2640, Pe\~{n}alol\'{e}n, Santiago, Chile \\
$^{4}$Data Observatory Foundation, Chile\\
$^{5}$Instituto de Estudios Astrof\'isicos, Facultad de Ingenier\'ia y Ciencias, Universidad Diego Portales, Av. Ej\'ercito Libertador 441, Santiago, Chile\\
$^{6}$European Southern Observatory, Alonso de Cordova 3107, Casilla 19001, Vitacura, Santiago, Chile\\
}

\date{Accepted XXX. Received YYY; in original form ZZZ}

\pubyear{2023}

\begin{document}
\label{firstpage}
\pagerange{\pageref{firstpage}--\pageref{lastpage}}
\maketitle

\begin{abstract}
  The $^{12}$CO rotational lines in protoplanetary discs are good tracers of the total spatial extension of the gas component, and potentially planet-disc interactions. We present ALMA long baseline observations of the $^{12}$CO(2-1) line of ten protoplanetary discs from the "Ophiuchus DIsc Survey Employing ALMA" (ODISEA) project, aiming to set constraints on the gas distribution of these sources. The position angle of the gaseous disc can be inferred for five sources using high-velocity channels, which trace the gas in the inner part of the disc. We compare the high-velocity PAs to the orientations inferred from the continuum, representative of the orientation over $\sim$ 53 to 256 au in these resolved discs. We find a significant difference in orientation for DoAr 44, which is evidence of a tilted inner disc. Eight discs show evidence of gas inside inner dust cavities or gaps, and the disc of ISO-Oph 196 is not detected in $^{12}$CO(2-1), except for the compact signal located inside its dust cavity. Our observations also point out a possible outflow in WLY 2-63.

\end{abstract}




\section{Introduction}

Advances in observational techniques and instruments have revealed a vast diversity of substructures in the components of circumstellar discs, such as spirals, concentric rings, gaps, and shadows that can be an indicator of warped morphologies \citep[e.g.][]{Casassus2013, Marino2015, Perez2016, Andrews2018}. However, the origin of all of these substructures is still unclear, and a link with embedded protoplanetary systems remains to be established.\\
In this context, a promising research avenue has been proposed through the observations of the gaseous components using CO tracers, which can reveal perturbations in the kinematics of the disc produced by the presence of a forming planet \citep{Perez2015, Perez2018}. Indeed, tentative detections of protoplanet candidates have been reported using the kinematics of the disc sampled in CO lines \citep[e.g.][]{Teague2018, Pinte2019}. Besides providing hints about planet-disc interactions, observations of CO isotopologues allow us to study fundamental physical properties in discs, such as the total gas mass, CO gas depletion, and temperature structures \citep[e.g.][]{Williams2014, VanDerMarel2016, Krijt2020, Zhang2021}.\\
After H$_{2}$, CO is the most abundant molecule in a protoplanetary disc, and its millimeter wavelength rotational lines are the strongest that are observable; therefore, CO is the most widely used gas tracer. Demographic studies would provide crucial information on the connection between the features seen in  CO lines with disc evolution \citep[e.g.][]{Ansdell2016, Pascucci2016, Grant2021, https://doi.org/10.48550/arxiv.2203.09930}. Regarding these studies, most of the research has been for the continuum. One major demographic study has been the "Ophiuchus DIsc Survey Employing ALMA" (ODISEA) project \citep{Cieza2019}. It has focused on studying the whole group of objects identified by the Spitzer Legacy project “Cores to discs” \citep{Evans_2009} located in the Ophiuchus star-forming region. The latter has the most significant number of discs of all nearby star-forming regions, at just 140 pc from the Sun \citep{Canovas2019}. ODISEA aims to study both the components of dust and gas of the Ophiuchus' discs, which has been done through the analysis of ALMA observations of continuum emission and the $^{12}$CO(2-1) line \citep{Cieza2019}. Interesting results from this project have already been published in recent years, for instance, \citet{Williams2019} provided measurements for the dust mass of 279 discs in Ophiuchus, \citet{Zurlo2020} studied properties of ODISEA discs with stellar multiplicity using NIR observations and detected 21 new multiple systems, and \citet{Gonzalez-Ruilova2020} studied the properties of the binary system ISO-Oph 2, finding a possible bridge of gas connecting primary and secondary disc.\\
A follow-up of 10 of the 15 brightest ODISEA sources at 1.3 mm was carried out in 2019, which consisted of ALMA band 6 long-baseline observations (1.3 mm/230 GHz) of continuum emission and the $^{12}$CO(2-1) molecular line. The continuum observations were presented in \citet{Cieza2021}, however, the line data from this follow-up (excluding one source, ISO-Oph 2) has remained unexplored.\\
Here we focus on the $^{12}$CO line of the long-baseline ODISEA data \citep{Cieza2021} and provide new observational constraints on the gas distribution of this sample of protoplanetary discs. In five cases, we measure the position angle using high-velocity channels that trace the gas near the star, and compare the orientation of the CO gas with that of the continuum. In addition, we measure the gas disc radius in the two cases devoid of diffuse $^{12}$CO(2-1) screens. Throughout this paper, we make use of the 1.3 mm continuum data from the same data set previously presented by \citet{Cieza2021}, and we also consider the SED classes indicated in that work for each source that are based on the definitions of \citet{Williams2011}. In this classification scheme, we consider Class I, Flat spectrum, Class II, and Class III young stellar objects. Besides, Class II objects are divided into three types, full discs, pre-transition discs (PTD), and transition discs (TD), depending on the shape of their SED between 2 and 20 $\mu$m.\\
This paper is organized as follows. In Section\,\ref{section:Observations} we briefly describe the data used; in Sec.\,\ref{sec:Analysis} we explain the method used to measure gas radii and position angles; we comment on each system and previous observations in Sec.\,\ref{sec:Sources}; and finally, in Sec.\,\ref{sec:Conclusions} we discuss the main results of this work.

\begin{table*}
	\centering
	\caption{Observing log.}
	\label{tab:table1}
	\begin{tabular}{lccccr} 
		\hline
		\hline
		Name & Date & Time on source  & Number of antennas & Baselines & Precipitable water vapour\\
		& & (sec) & & (m) & (mm)\\
		\hline
		ISO-Oph 54 & 2019-07-12 & 3183 & 41 & 111.2 to 12644.7 & 1.2\\
		WLY\,2-63 & 2019-06-24 & 3058 & 48 & 83.1 to 16196.3 & 0.4\\
		ISO-Oph 37 & 2019-07-08 & 3223 & 44 & 149.1 to 13894.4 & 1.2\\
		ISO-Oph 17 & 2019-06-19 & 3069 & 44 & 83.1 to 16196.3 & 1.1\\
		DoAr\,44 & 2019-07-11 & 3258 & 44 & 111.2 to 12644.7 & 1.5\\
		DoAr\,44 & 2019-07-13 & 3256 & 43 & 111.2 to 12644.7 & 1.1\\
		WSB 82 & 2019-06-05 & 3162 & 45 & 83.1 to 15238.4 & 1.0\\
		ISO-Oph 2 & 2019-06-12 & 2119 & 45 & 83.1 to 16196.3 & 1.2\\
		ISO-Oph 2 & 2019-06-21 & 3089 & 46 & 83.1 to 16196.3 & 0.9\\
		ISO-Oph 196 & 2019-07-11 & 3204 & 44 & 111.2 to 12644.7 & 1.4\\
		SR\,24S & 2019-07-12 & 3199 & 43 & 111.2 to 12644.7 & 1.2\\
		RXJ1633.9-2442 & 2019-06-24 & 3142 & 48 & 83.1 to 16196.3 & 0.4\\
		\hline
	\end{tabular}
\end{table*}

\begin{table*}
	\centering
	\caption{Properties of the $^{12}$CO Datacubes.}
	\label{tab:table2}
	\begin{tabular}{lccccccr} 
		\hline
		\hline
		Source & Synthesized beam & Rms Noise & Peak & Peak & Flux & robust & uvtaper\\
		 & & (Jy beam$^{-1}$) & (Jy beam$^{-1}$) & S/N & (Jy\,km\,s$^{-1}$) & & \\
		\hline
		ISO-Oph 54 & 0.11" $\times$ 0.13" & 0.0013 & 0.011 & 6.88 & 0.16 & 2 & 0.08" $\times$ 0.08"\\
		WLY\,2-63 & 0.14" $\times$ 0.15" & 0.0036 & 0.053 & 13.41 & 4.24 & 2 & 0.13" $\times$ 0.13"\\
		ISO-Oph 37 & 0.14" $\times$ 0.16" & 0.0019 & 0.062 & 32.89 & 1.93 & 2 & 0.13" $\times$ 0.13"\\
		ISO-Oph 17 & 0.11" $\times$ 0.12" & 0.0020 & 0.042 & 7.89 & 1.47 & 2 & 0.1" $\times$ 0.1"\\
		DoAr\,44 & 0.12" $\times$ 0.13" & 0.00096 & 0.030 & 28.49 & 2.12 & 2 & 0.1" $\times$ 0.1"\\
		WSB 82 & 0.14" $\times$ 0.15" & 0.0022 & 0.047 & 22.19 & 1.44 & 2 & 0.13" $\times$ 0.13"\\
		ISO-Oph 2 & 0.11" $\times$ 0.12" & 0.0018 & 0.027 & 12.33 & 0.69 & 2 & 0.1" $\times$ 0.1"\\
		ISO-Oph 196 & 0.12" $\times$ 0.14" & 0.0017 & 0.011 & 6.79 & 0.029 & 2 & 0.1" $\times$ 0.1"\\
		SR\,24S & 0.12" $\times$ 0.14" & 0.0016 & 0.048 & 30.84 & 1.52 & 2 & 0.1" $\times$ 0.1"\\
		RXJ1633.9-2442 & 0.13" $\times$ 0.14" & 0.0017 & 0.037 & 20.16 & 2.24 & 2 & 0.13" $\times$ 0.13"\\
		\hline
	\end{tabular}
\end{table*}

\section{Observations}
\label{section:Observations}
We present ALMA band 6 (1.3 mm) data of ten sources that were taken as part of the project $2018.1.00028.$S. The observations were performed in 2019 between June and July. ISO\,Oph\,2 data from both continuum and the $^{12}$CO line were extensively analysed in \citet{Gonzalez-Ruilova2020}.\\
The correlator setup was chosen to maximize the continuum sensitivity, therefore, only one line was incorporated. The observations were configured in Time Division Mode and taken in four spectral windows. Three windows were centered at 217, 219, and 233 GHz for the continuum. Further, the fourth spectral window was centered on the $^{12}$CO J $=$ 2$-$1 line (230.538 GHz) and had a 1.5~km~s$^{-1}$ velocity resolution.\\
All data were first calibrated using the ALMA pipeline (for more detailed information about this data set, see \citet{Cieza2021} and Table\,\ref{tab:table1}). Continuum subtraction was applied in all cases, and the line imaging process was performed using the {\sc tclean} task in the Common Astronomy Software Applications package ({\sc CASA}, \citet{Mcmullin2007}) v.6.1.2, with Briggs weighting and a robust parameter of 2. Tapering was also applied in all cases to improve the sensitivity of the images. The details of each data cube are summarized in Table \ref{tab:table2}, and the basic properties of each source are summarized in Table \ref{tab:table3}.\\ 
Since self-calibration improved the peak SNR for the continuum by only 10-30\% \citep[][]{Cieza2021}, we decided not to apply the gain calibration tables to the line data. We did not explore Keplerian masking because not all the sources display clear keplerian features, and traditional masking with {\sc tclean} yielded satisfactory results for the sources free of diffuse emission. We explored the possibility of combining the long-baseline data with lower-resolution observations from the first data set of the ODISEA survey (PID = 2016.1.00545.S). However, we found that, given the short on-source time (45 s) of the low-resolution data,  the combined data products were noisier than those produced with the long-baseline data by themselves. Therefore,  we chose to focus on the long-baseline data alone.\\
We generated moment 0 and moment 1 maps for the sources less affected by contamination from the cloud using the {\sc immoments CASA} task, and generated the maps with all the spectral channels where the source is conspicuous. In addition, we measured fluxes using region statistics in {\sc viewer} within {\sc CASA}.

\begin{table*}
	\centering
	\caption{Basic properties of each source. (0): Inclination from Cieza et al. 2021. (1): Position angle measured for the $^{12}$CO(2-1) emission (this work). (2): Position angle of the dust from Cieza et al. 2021. (3) and (4): Coordinates from Gaia, excluding ISO-Oph 54, WLY\,2-63, ISO-Oph 37 and ISO-Oph 17 (coordinates from the 2MASS All-sky catalog).}
	\label{tab:table3}
	\begin{tabular}{lcccccr} 
		\hline
		\hline
		Name & SED Class & \textit{i} & Gas P.A. & Continuum P.A. & Ra & Dec\\
		& & (deg) & (deg)& (deg) & (deg) & (deg)\\
		& & (0) &(1) & (2) & (3) & (4)\\
		\hline
		ISO-Oph 54 & I & 32.5 $\pm$ 0.4 & ... & 160 $\pm$ 2.5 & 246.6685833 & -24.4539722\\
		WLY\,2-63 & FS & 46.2 $\pm$ 1.2 & ... & 149 $\pm$ 5.0 & 247.8985416 & -24.0249722\\
		ISO-Oph 37 & FS & 72.4 $\pm$ 0.2 & ... & 49 $\pm$ 0.6 & 246.5982083 & -24.4109444\\
		ISO-Oph 17 & FS & 42.4 $\pm$ 0.7 & 132.9 $\pm$ 2.5 & 131 $\pm$ 0.7 & 246.5430000 & -24.3485000\\
		DoAr\,44 & II/PTD & 21.8 $\pm$ 0.9 & 94.1 $\pm$ 8.6 & 60 $\pm$ 2.7 & 247.8894038 & -24.4604259\\
		WSB 82 & II/Full & 61.2 $\pm$ 0.5 & 165.9 $\pm$ 5.2 & 173 $\pm$ 1.0 & 249.9393177 & -24.0345083\\
		ISO-Oph 2 & II/PTD & 37.6 $\pm$ 0.8 & ... & 0.4 $\pm$ 1.4 & 246.4088384 & -24.3768303\\
		ISO-Oph 196 & II/Full & 22.3 $\pm$ 1.4 & ... & 132 $\pm$ 7.1 & 247.0687710 & -24.6162380\\
		SR\,24S & II/PTD & 47.3 $\pm$ 3.1 & 26.1 $\pm$ 2.2 & 28 $\pm$ 1.2 & 246.7437779 & -24.7603309\\
		RXJ1633.9-2442 & II/TD & 47.9 $\pm$ 1.1 & 79.4 $\pm$ 4.8 & 77 $\pm$ 1.4 & 248.4817058 & -24.7014913\\
		\hline
	\end{tabular}\\

\end{table*}

\section{Analysis}
\label{sec:Analysis}

\subsection{Measurements of gas disc radii}

\begin{figure*}
\includegraphics[width=0.48\textwidth]{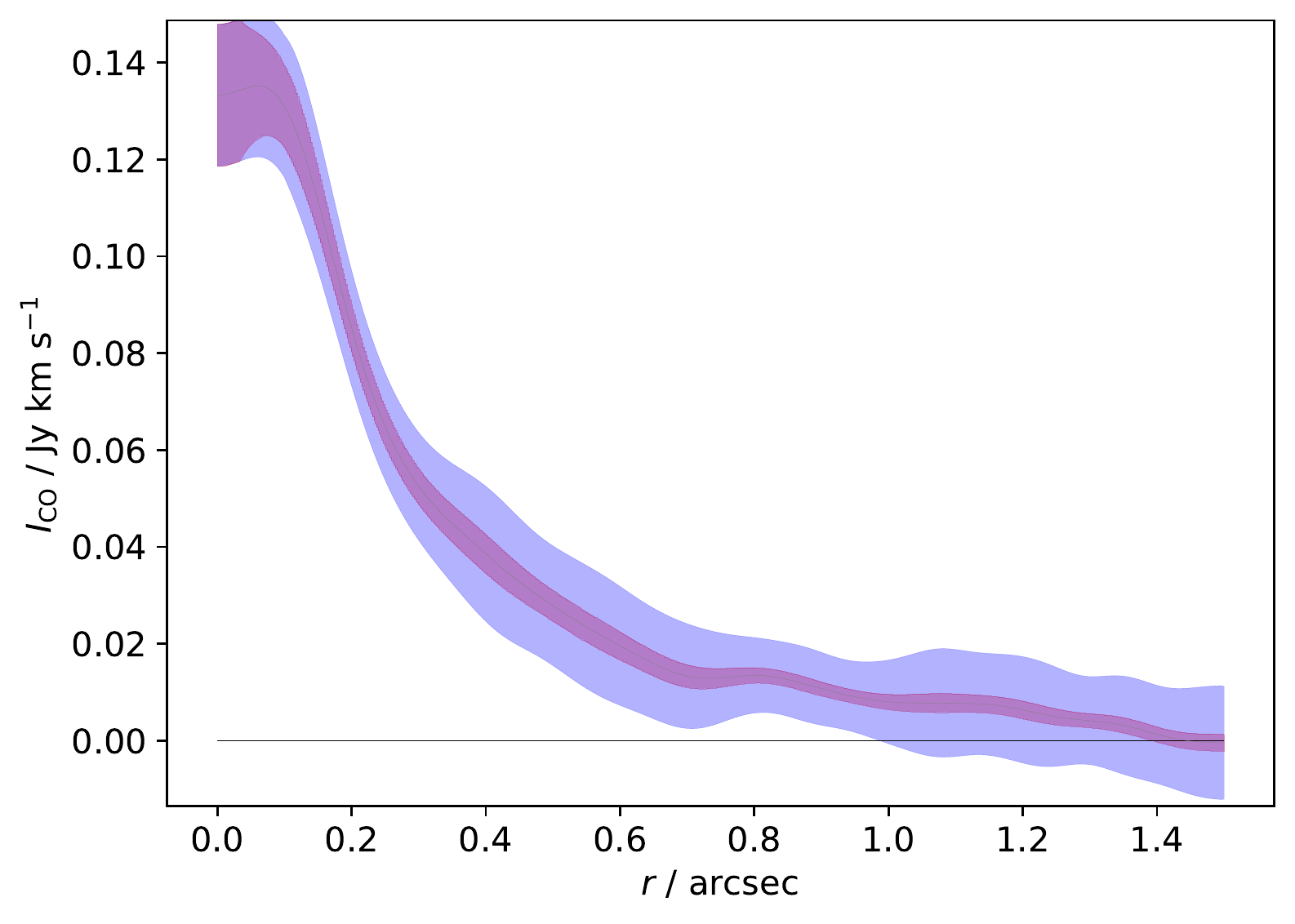}
\includegraphics[width=0.48\textwidth]{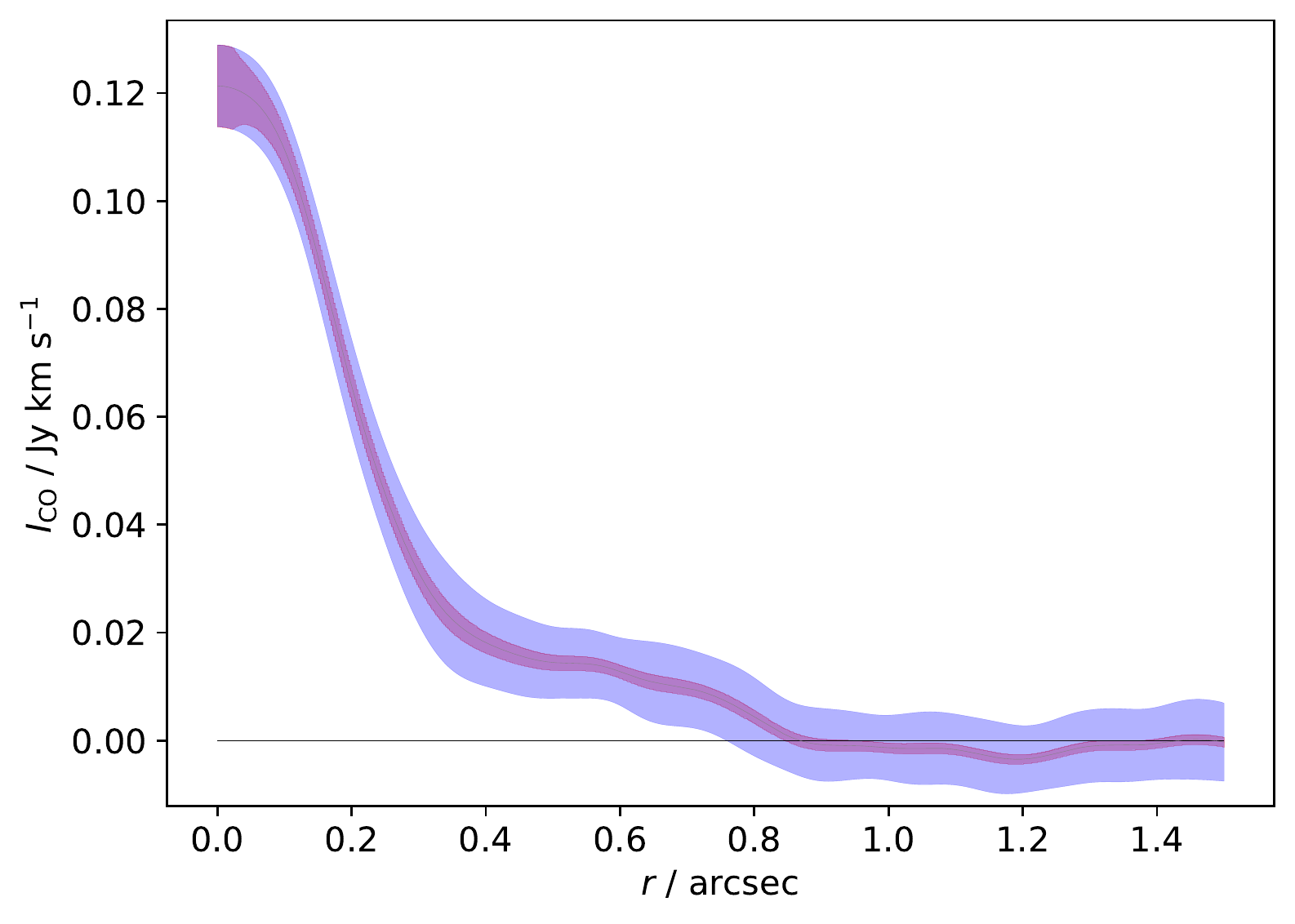}
\caption{Deprojected radial brightness profiles of the $^{12}$CO(2-1) emission for: RXJ1633.9-2442 on the left, and DoAr\,44 on the right side. We plot the azimuthally averaged radial profiles, along with the error on the mean and dispersion as shaded areas (both $\pm1\sigma$). }
\label{fig:radial-profiles-gas}
\end{figure*}

We measured the gas size for two discs in our sample without significant cloud contamination: RXJ1633.9-2442 and DoAr\,44. Both measurements were performed using radial profiles in Fig.\,\ref{fig:radial-profiles-gas}. We deprojected moment 0 images using inclinations and position angles derived for the continuum \citep{Cieza2021}. After that, extracted azimuthally averaged radial profiles of the $^{12}$CO(2-1) emission. We considered the following definition of integrated disc flux:
\begin{equation}
    F(r) = 2\pi \int_{0}^{r}I(s) s ds 
\end{equation}
where $s$ is the radial coordinate projected on the sky, and $I(s)$ the observed intensity; we calculated the gas disc radius $R_{gas}$ as the radius containing 90\% of the total flux $F(\infty)$; besides, the uncertainties were calculated considering the radius of the major axis of the synthesized beam. We obtained $R_{gas}$ $161.0 \pm 9.8$ au and $80.9 \pm 9.1$ au for RXJ1633.9-2442 and DoAr\,44 respectively; we also compared the $R_{gas}$ with the dust radius $R_{90\%}$ measured in \citet{Cieza2021} and we obtained $R_{gas}/R_{dust}$: \textbf{$\sim$} 3.0 and 1.4 respectively. 

\subsection{Measurements of disc position angles}
We measured the position angle (P.A.) using high-velocity channels to probe the orientation of internal gas discs. Besides, we assume that the P.A. of the outer disc is the P.A. of the mm continuum, previously reported in Cieza et al. 2021. To compute the position angle, we performed a two-dimensional fitting of an elliptical Gaussian to the emission of two high-velocity channels of each data set using the {\sc imfit} task within {\sc CASA}. This fit was done for the most redshifted and blueshifted emission of the data cubes, here we select channels of significant emission above the noise level, specifically, with peak S/N (ratio between peak and rms) greater than six. After getting a centroid for each channel, a line was drawn to connect both centroids. This line corresponds to the disc position angle and is measured in the direction east of north. The results of this method are shown in Table\,\ref{tab:table3}, and we present a comparison between the P.A. measured in this work and the one measured for the dust emission in \citet{Cieza2021}  in Sec.\,\ref{sec:Sources}. Since we used $\alpha=\arctan{\frac{\Delta\rm{ra}}{\Delta\rm{dec}}}$, we estimated the uncertainty of the P.A. by propagating errors, starting with the uncertainty of the position of each centroid. We were able to apply this method for five sources of our sample (we refer to this sample as "discs with clear Keplerian features") because in the other five cases, we either identified emission in few channels or the morphology of the $^{12}$CO had peculiar features (we refer to this sample as "discs without clear Keplerian features").

\section{Discussion}
\label{sec:Sources}

In the following section, we discuss the $^{12}$CO($J=$2$-$1) molecular line observations for each source. In five cases, we compare the P.A. measured for the gas component in this work with the P.A. of the dust previously measured in \citet{Cieza2021}. For a detailed description of the fundamental parameters of each star's SED and previous observations of the dust emission, we recommend seeing \citet{Cieza2021} and references therein.

\subsection{Individual sources: discs with clear Keplerian features}

\subsubsection{DoAr\,44}
\begin{figure*}
\includegraphics[width=0.498\textwidth]{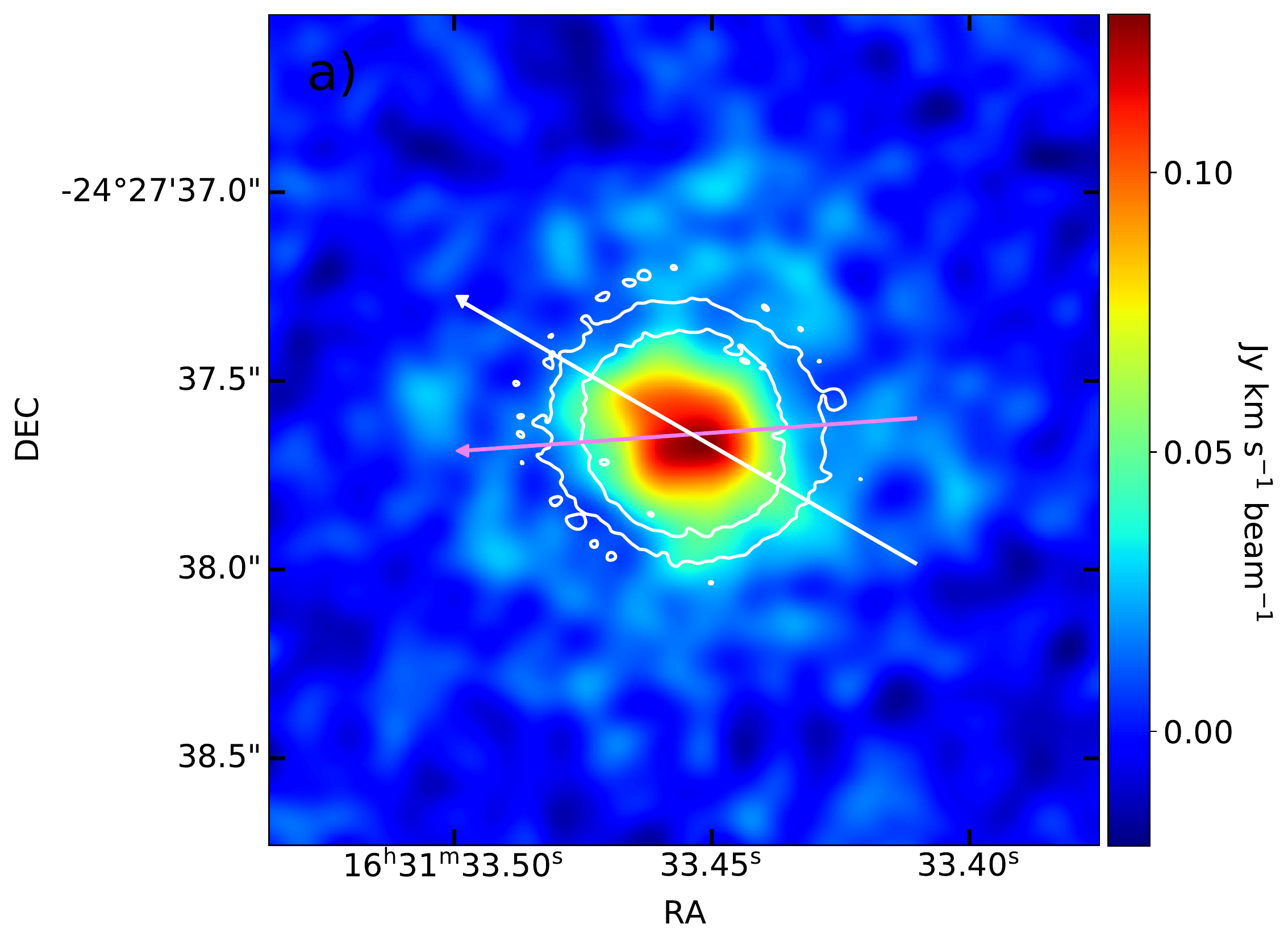}
\includegraphics[width=0.439\textwidth]{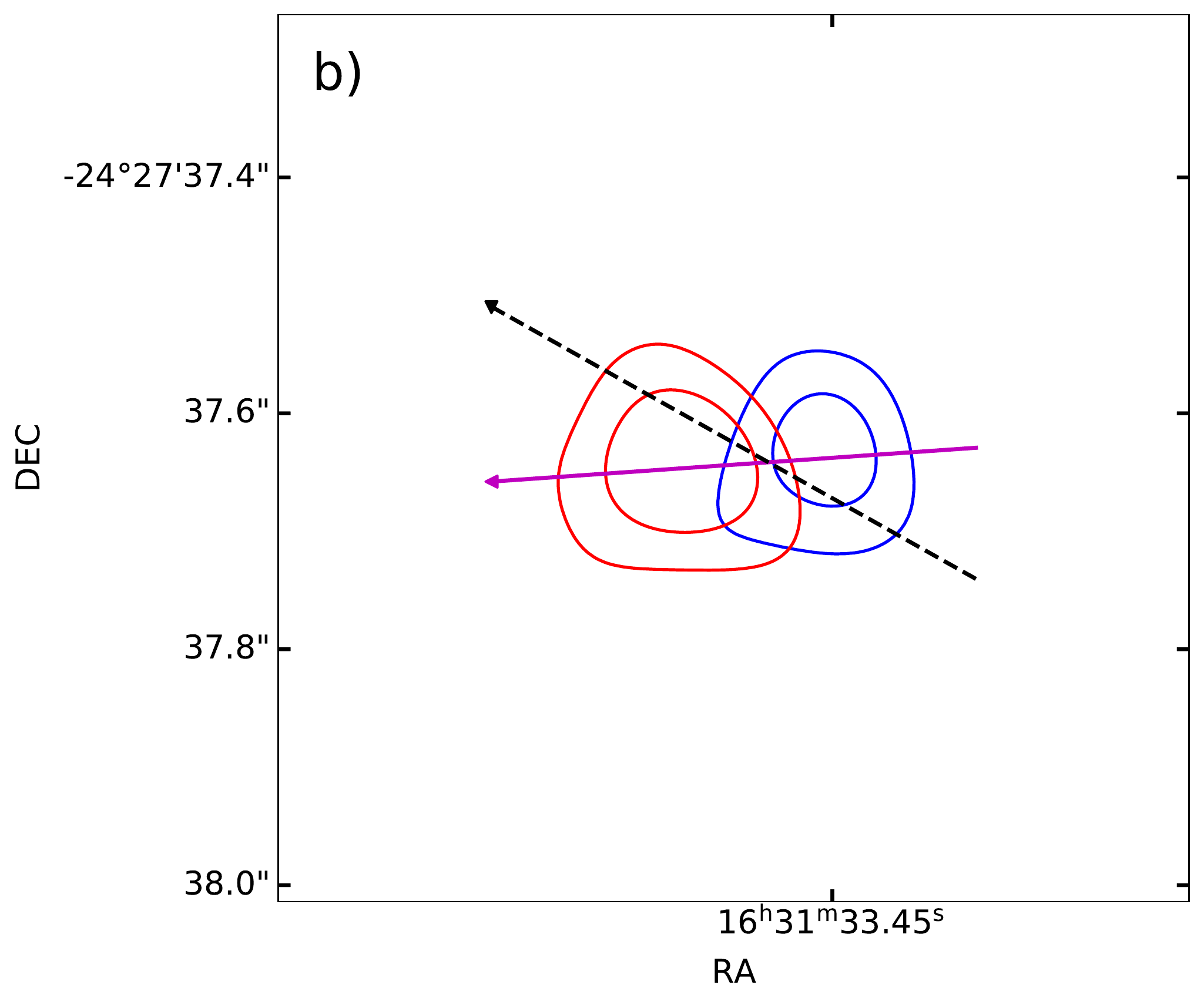}
\includegraphics[width=7.8cm]{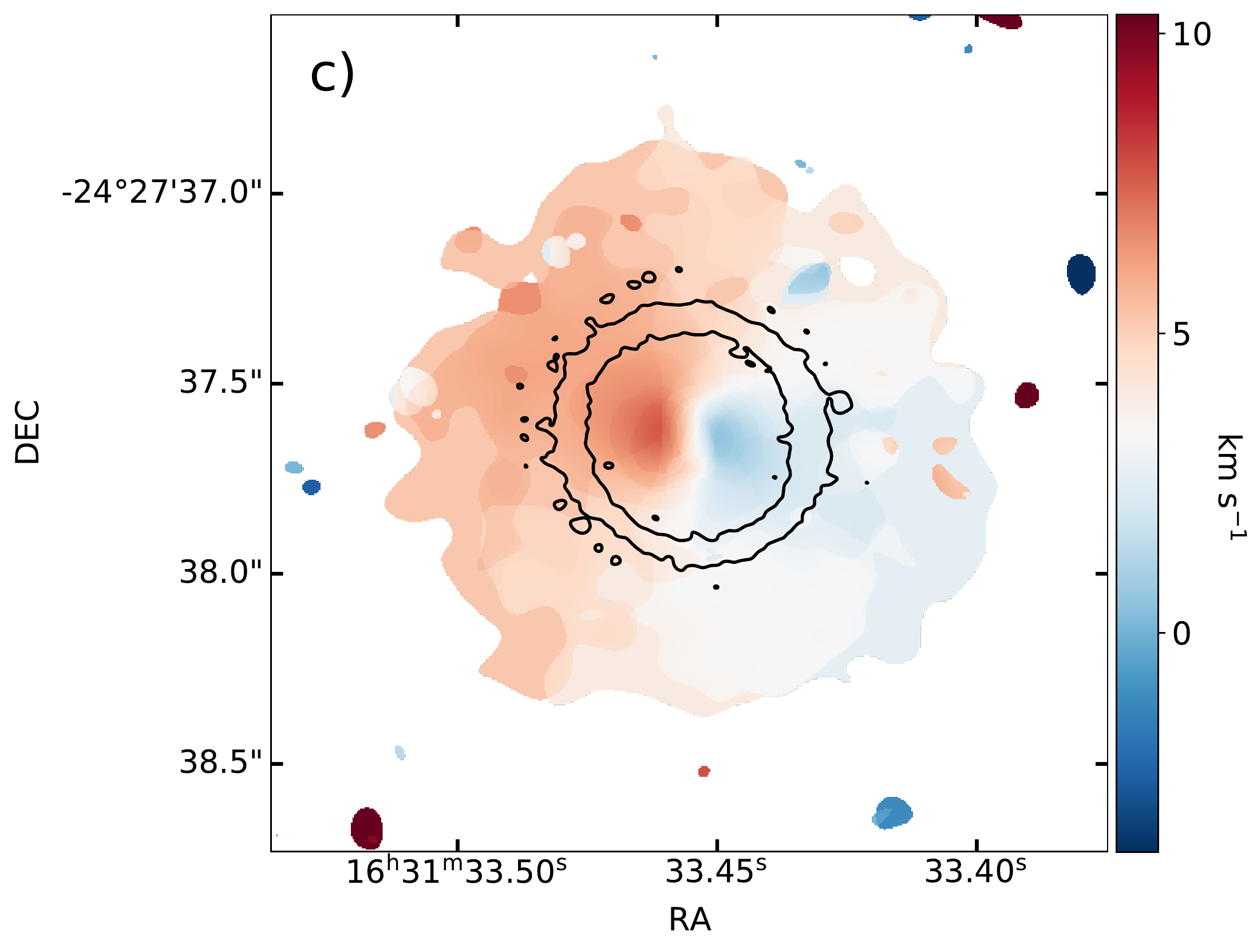}

\caption{(a) Moment 0 map of the $^{12}$CO line for DoAr\,44, the integrated velocities range from -3.65 to 10.32~km~s$^{-1}$, contours of continuum emission at the 5$\sigma$ level are indicated in white. The white arrow symbolizes the dust position angle, while the magenta arrow symbolizes the position angle from this work. (b) Illustration of the P.A. measurement, contours of the most redshifted $^{12}$CO emission are drawn in red while contours of the most blueshifted emission are drawn in blue, contours represent 3 and 6 mJy beam$^{-1}$. The magenta arrow symbolizes the position angle of the gas, and the black dashed line the position angle of the dust. (c) Moment 1 map of $^{12}$CO emission for DoAr\,44, the integrated velocities range from -3.65 to 10.32~km~s$^{-1}$.}
\label{fig:doarPA}
\end{figure*}

\begin{figure*}
\centering
    \includegraphics[width=15.7cm]{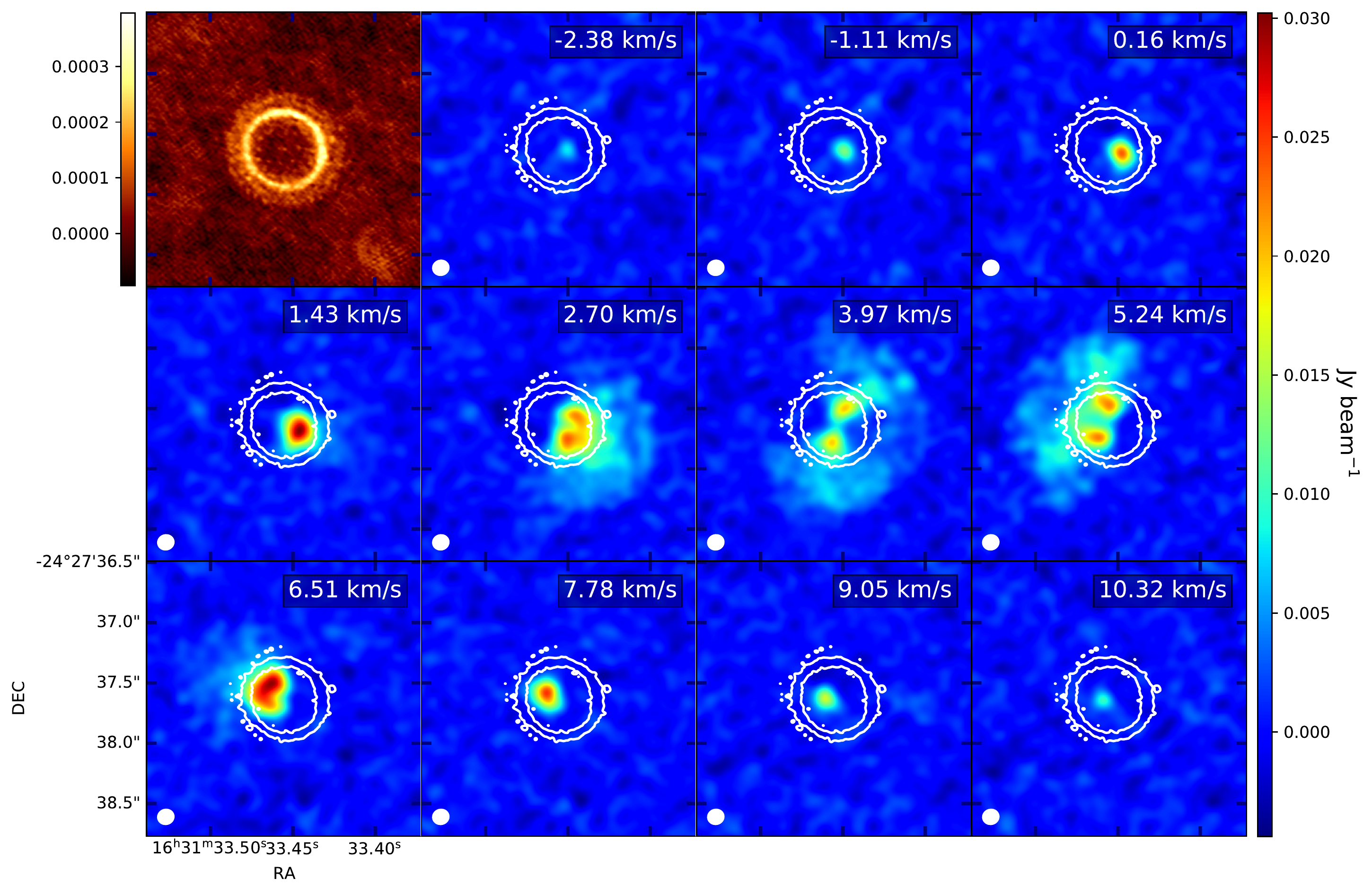}
    \caption{Observed channel maps of $\protect {}^{12}$CO(2-1) of DoAr\,44. An image of the 1.3 mm  continuum is shown at the top and to the left (continuum map from~\protect\citet{Cieza2021}). Contours of 1.3 mm continuum emission at the 5$\sigma$ level are indicated in white, and the white ellipses represent the synthesized beam. For the beam size, we refer to Table \ref{tab:table2}. Both color bars have units of Jy~beam$^{-1}$.}
    \label{fig:my_labelDO}
\end{figure*}

This source is also known as WSB 72 and HBC 268. It has a pre-transition disc SED. A large cavity has been observed in the submillimeter by \citet{VanDerMarel2016} and \citet{Cieza2021}, and resolved in the near IR by \citet{Avenhaus2018}. Besides, the observations of \citet{Avenhaus2018} showed shadows in the outer disc, and a central warp of the inner disc was proposed to explain those features \citep[][]{Casassus2018}.\\
When comparing the P.A. measured for the gas and the one measured for the continuum emission, there is a significant difference of 34.1 $\pm$ 9.0 degrees (Fig.\,\ref{fig:doarPA}). Therefore, our long-baseline observations suggest a difference in orientation between the inner gas and the mm-dust. The separation of both centroids in Fig.\,\ref{fig:doarPA} b) is $\sim$\,0,17", i.e. a distance of $\sim$\,24 au assuming a distance of 146 pc. Moreover, the moment 1 map in Fig.\,\ref{fig:doarPA} c) shows a twisted kinematic structure in the central area. This is clear evidence in favor of the warped morphology that has been proposed for this source. Further, in our channel maps in Fig.\,\ref{fig:my_labelDO} we identify gas emission inside the inner dust-depleted zone.

\subsubsection{RXJ1633.9-2442}
\begin{figure*}
\includegraphics[width=0.498\textwidth]{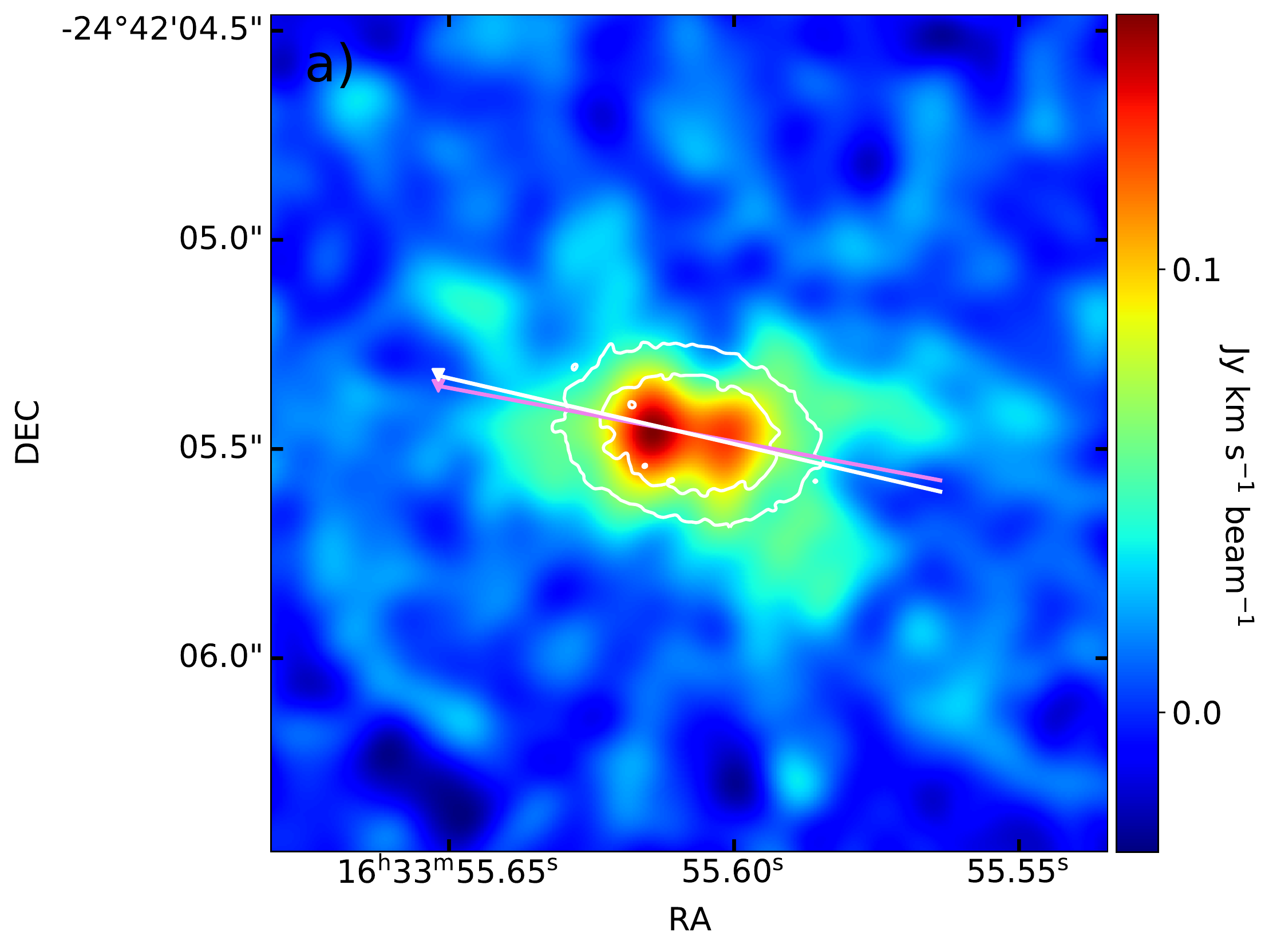}
\includegraphics[width=0.467\textwidth]{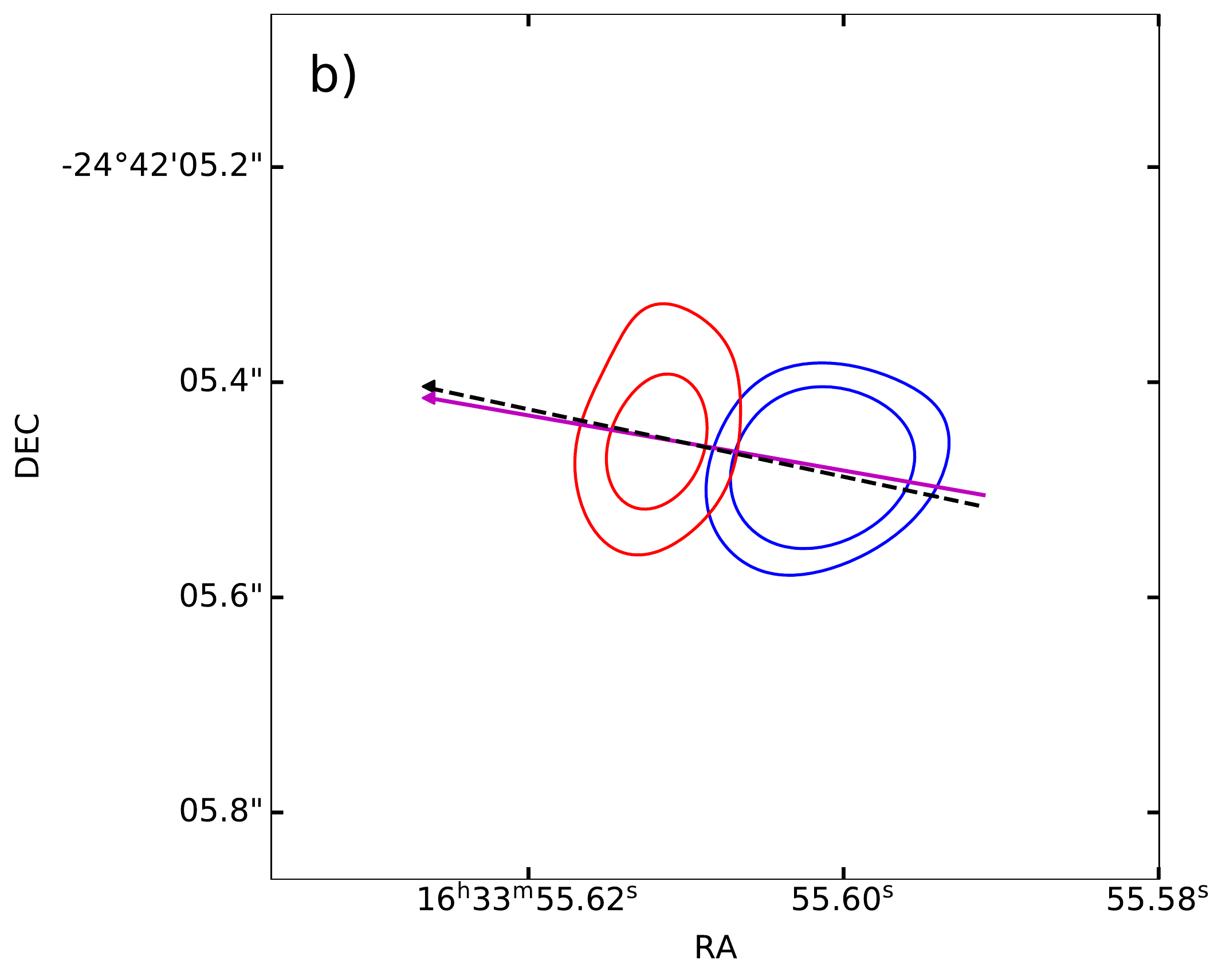}
\includegraphics[width=8.0cm]{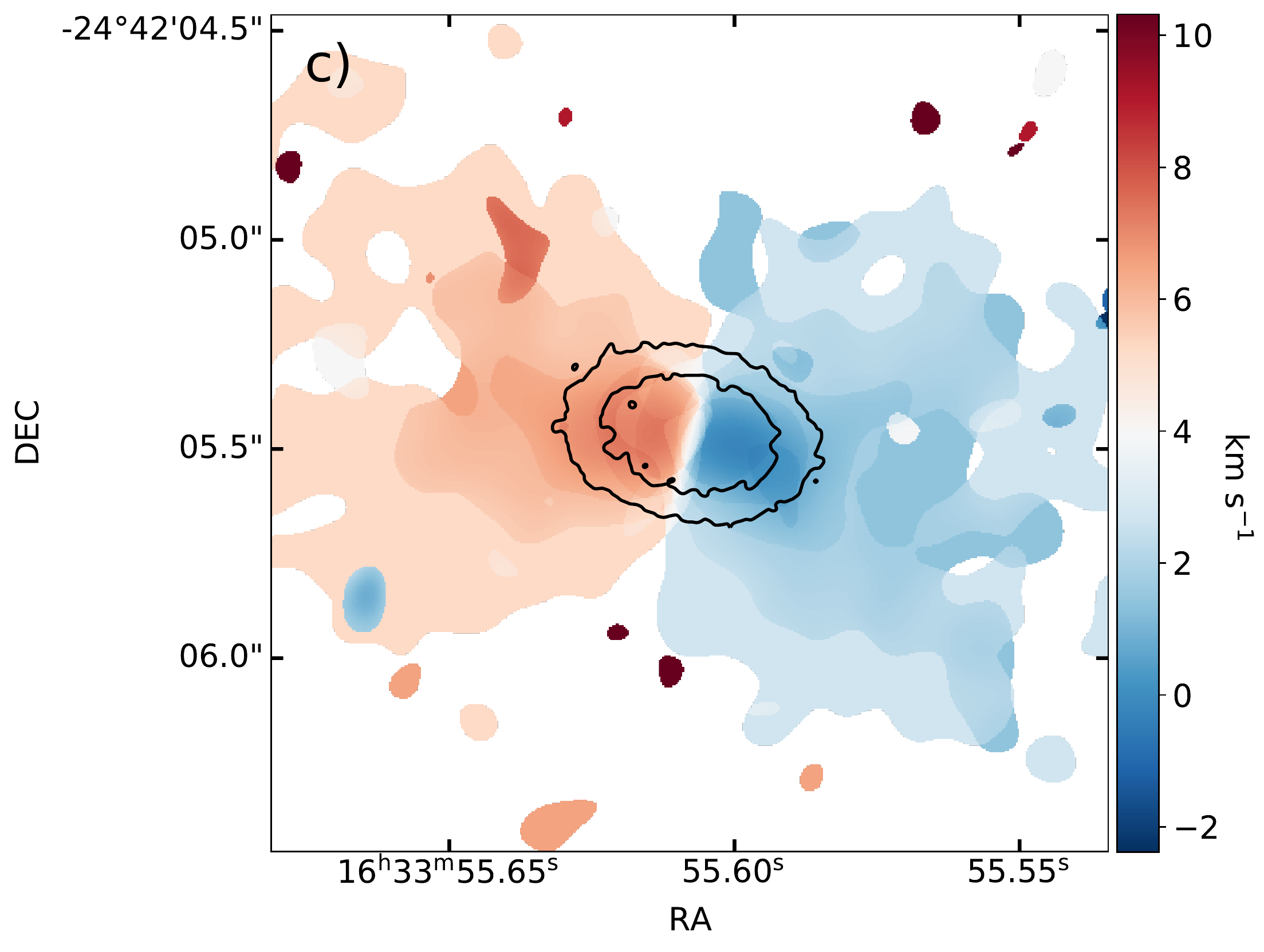}

\caption{(a) Moment 0 map of the $^{12}$CO line for RXJ1633.9-2442, the integrated velocities range from -2.38 to 10.32~km~s$^{-1}$, contours of continuum emission at the 9$\sigma$ level are indicated in white. (b) Contours represent 6 and 10 mJy beam$^{-1}$. (c) Moment 1 map of $^{12}$CO emission for RXJ1633.9-2442, the integrated velocities range from -2.38 to 10.32~km~s$^{-1}$. Annotations follow from Fig.\,\ref{fig:doarPA}.}
\label{fig:pa-rxj-633}
\end{figure*}

\begin{figure*}
\centering
    \includegraphics[width=15.7cm]{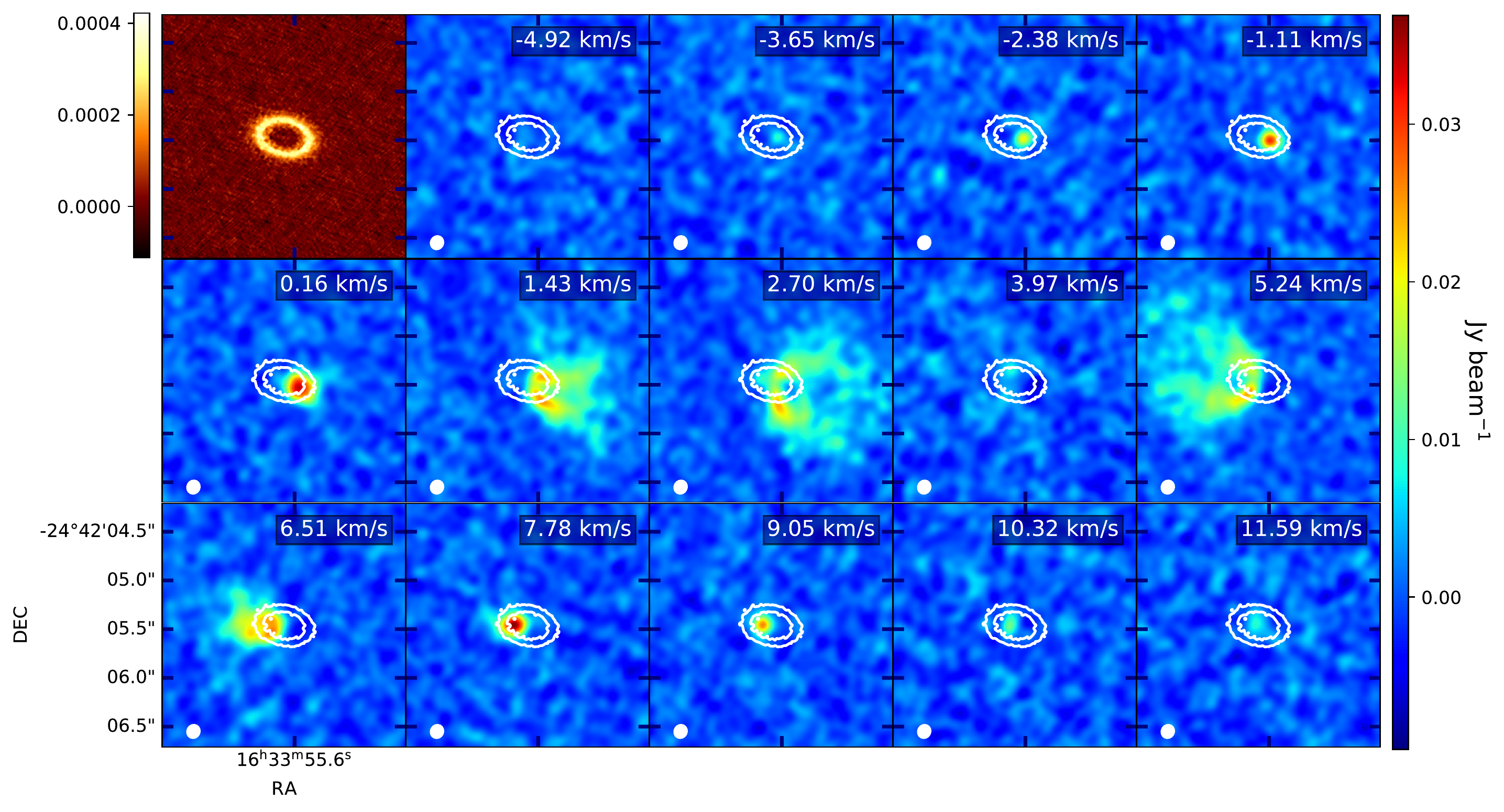}
    \caption{Observed channel maps of $\protect {}^{12}$CO(2-1) of RXJ1633.9-2442. Contours of 1.3 mm continuum emission at the 9$\sigma$ level are shown in white. Annotations follow from Fig.\,\ref{fig:my_labelDO}.}
    \label{fig:my_labelRX}
\end{figure*}

This source has a transition disc SED. \citet{Cieza2012} observed for the first time its inner cavity of dust with the Submillimeter Array (SMA) and the ODISEA data at 1.3 mm \citep[][]{Cieza2021} showed that the outer disc is composed of a narrow ring.\\
Fig.\,\ref{fig:pa-rxj-633} shows moment maps of $^{12}$CO emission for RXJ1633.9-2442, along with a comparison of P.A. for the gas and dust. No significant difference was
found between the P.A. measured for the gas and the P.A. of the dust. Our observations in Fig.\,\ref{fig:my_labelRX} display $^{12}$CO(2-1) emission within the inner gap and in the outer part of the disc showing a butterfly pattern. Besides, we notice mild contamination from the cloud material in all the channels in Fig.\,\ref{fig:my_labelRX}.

\subsubsection{WSB 82}
\begin{figure*}
\includegraphics[width=0.498\textwidth]{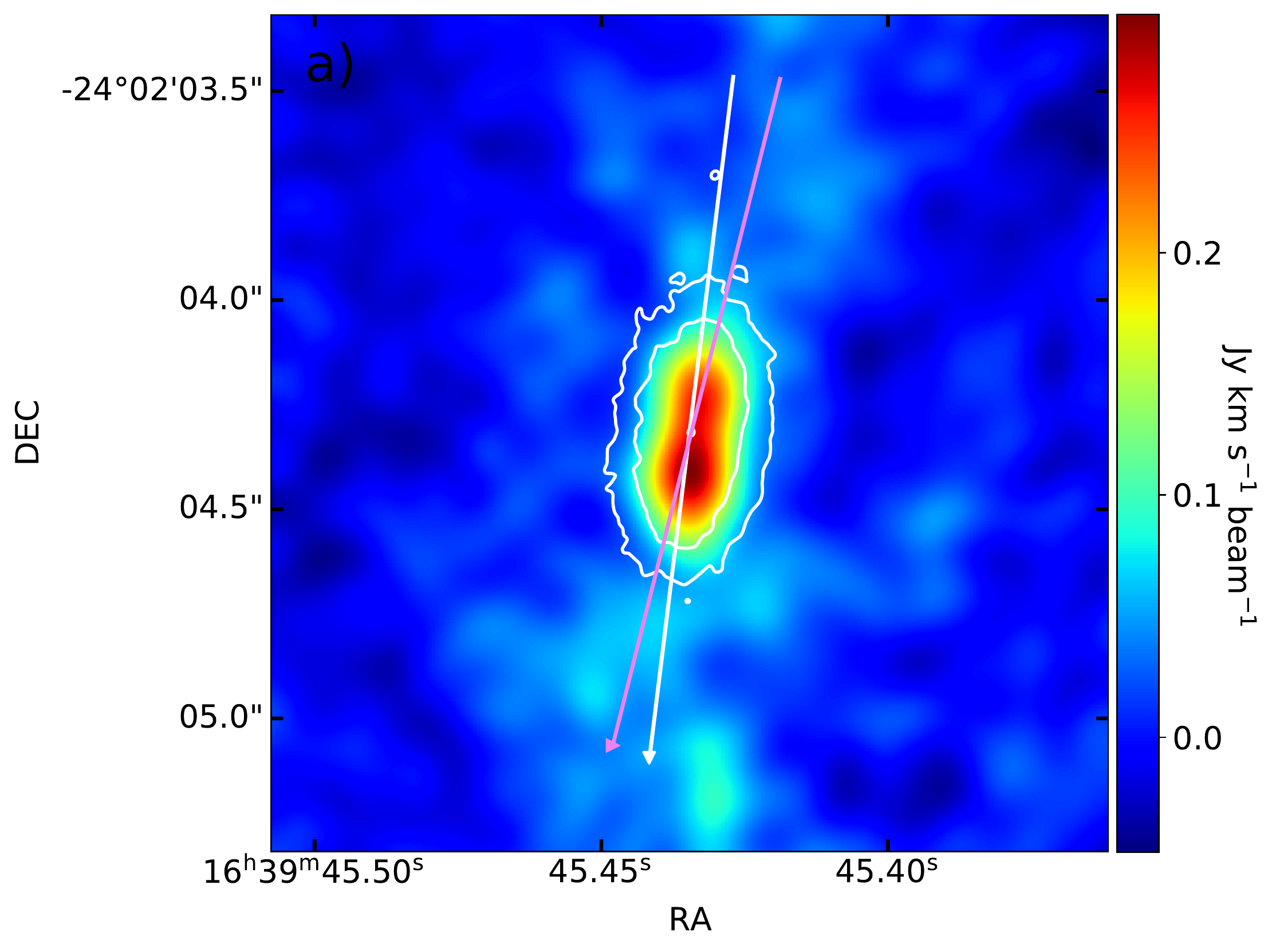} 
\includegraphics[width=0.445\textwidth]{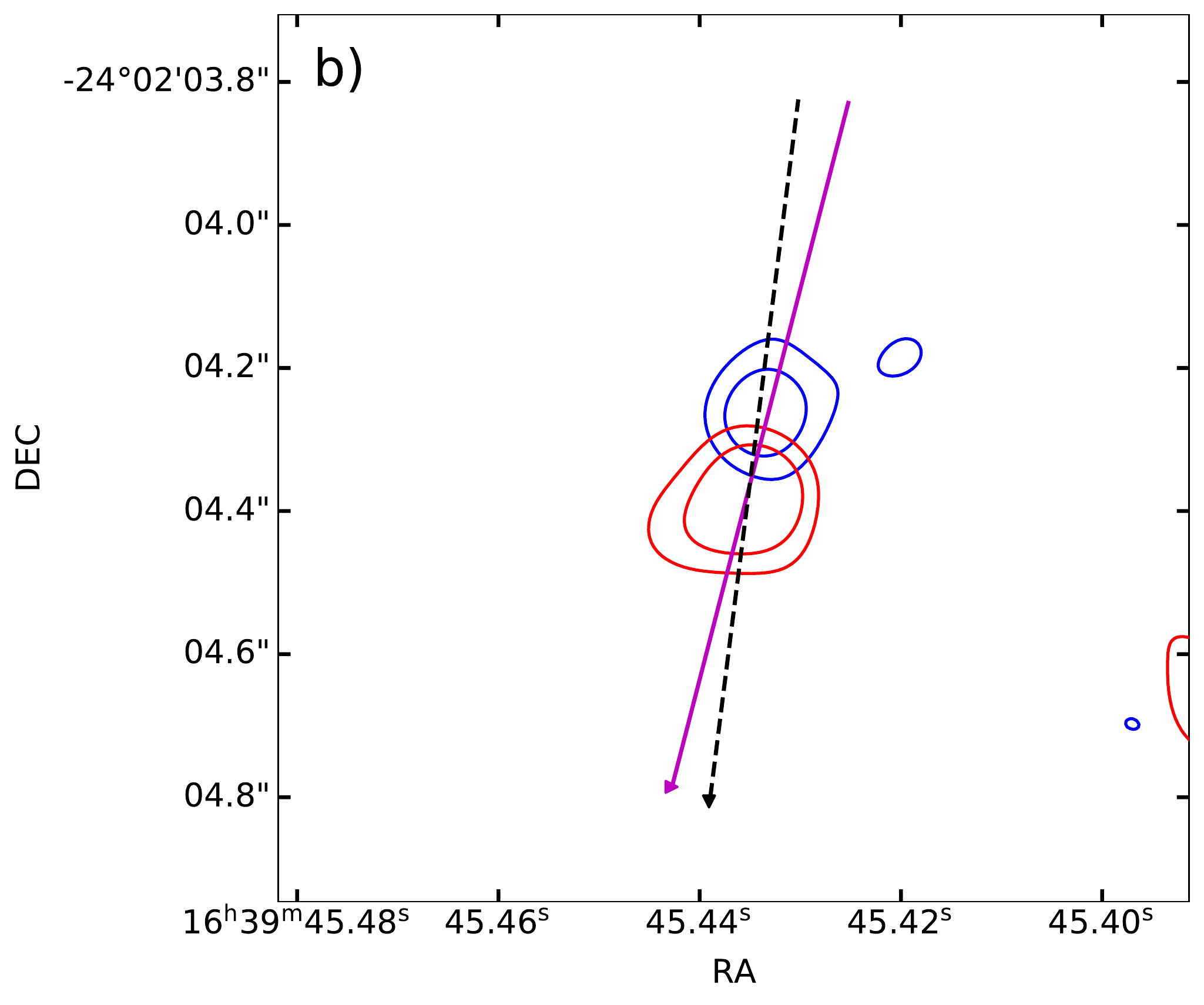}
\includegraphics[width=0.498\textwidth]{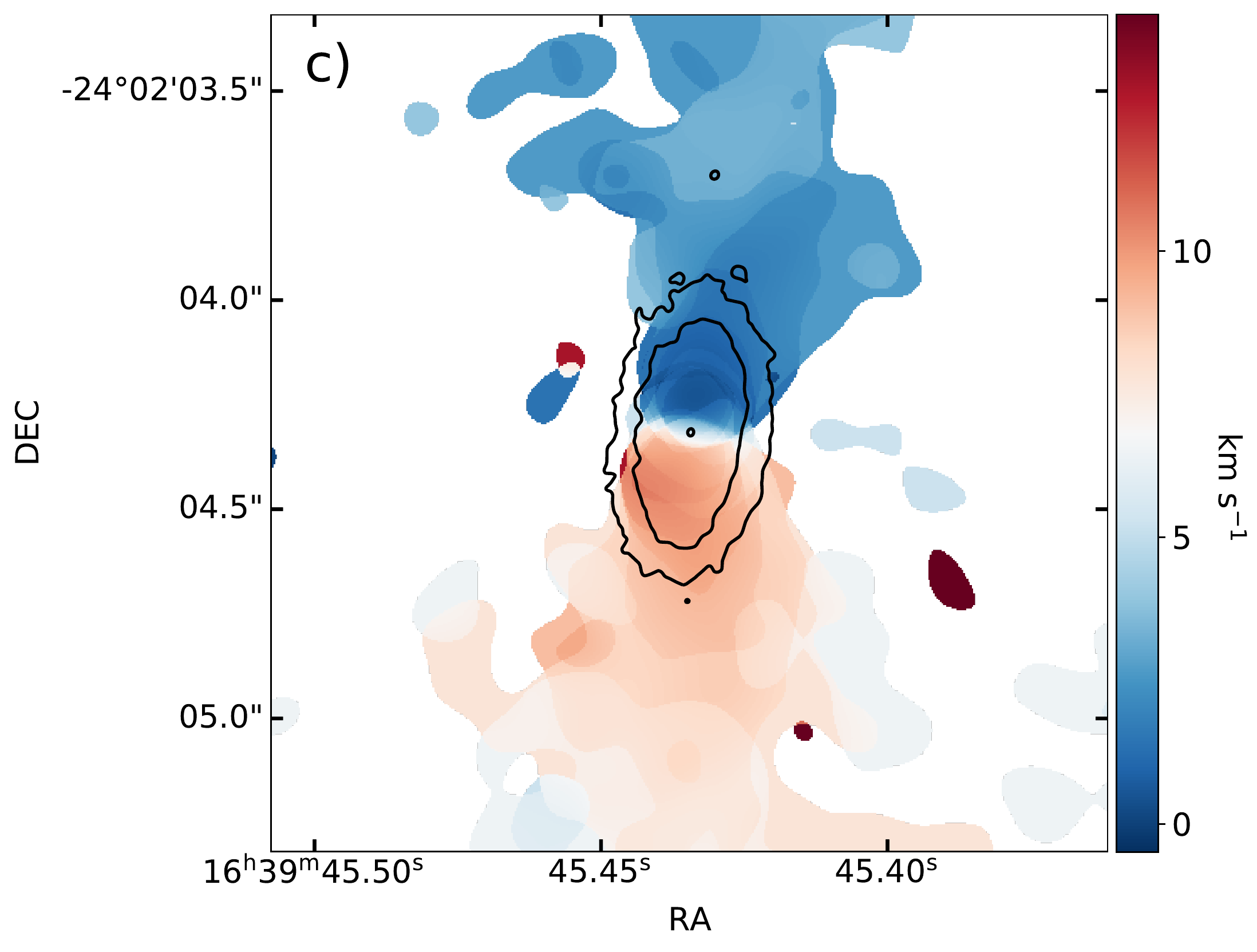} 
\caption{(a) Moment 0 map of the $^{12}$CO line for WSB\,82, the integrated velocities range from -4.92 to 14.13 km~s$^{-1}$, contours of continuum emission at the 9$\sigma$ level are indicated in white. (b) Contours represent 6.2 and 10 mJy/beam. (c) Moment 1 map of $^{12}$CO emission for WSB\,82, the integrated velocities range from -4.92 to 14.13~km~s$^{-1}$. Annotations follow from Fig.\,\ref{fig:doarPA}.}
\label{fig:pa-wsb82}
\end{figure*}

\begin{figure*}
\centering
    \includegraphics[width=15.7cm]{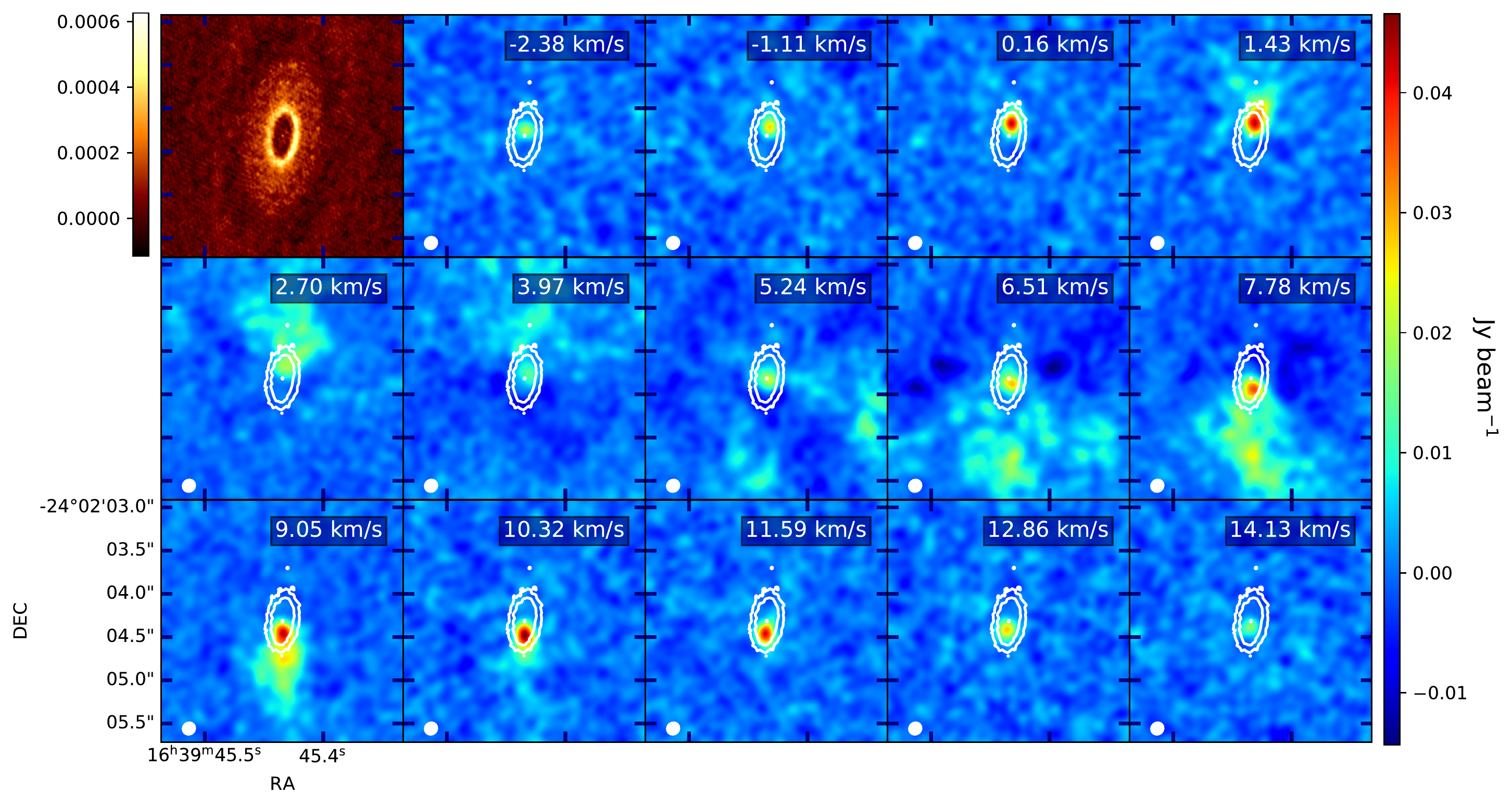}
    \caption{Observed channel maps of $\protect {}^{12}$CO(2-1) of WSB 82. Contours of 1.3 mm continuum emission at the 9$\sigma$ level are shown in white.  Annotations follow from Fig.\,\ref{fig:my_labelDO}.}
    \label{fig:my_label82}
\end{figure*}

This source is also known as IRAS 16367-2356 and has a Class II SED. \citet{Cox2017} and \citet{Cieza2021} identified a gap in its disc of dust, an outer disc with two pairs of gaps and rings at 1.3 mm, and an unresolved inner disc with a narrow ring at the edge of the inner gap.\\ Fig.\,\ref{fig:pa-wsb82} shows integrated intensity and weighted-velocity maps of $^{12}$CO emission for WSB\,82, together with an illustration indicating the P.A. for the gas and dust. No significant difference was found between the P.A. measured for the gas in this work and that previously measured for the dust. Channel maps in Fig.\,\ref{fig:my_label82} display $^{12}$CO(2-1) emission inside the inner dust cavity of the disc, in the outer disc, and gas emission widely distributed along the radial axis.  Besides, all the channels in Fig.\,\ref{fig:my_label82} exhibit moderate contamination because of emission from the cloud.

\subsubsection{ISO-Oph 17}
\begin{figure*}
\includegraphics[width=0.498\textwidth]{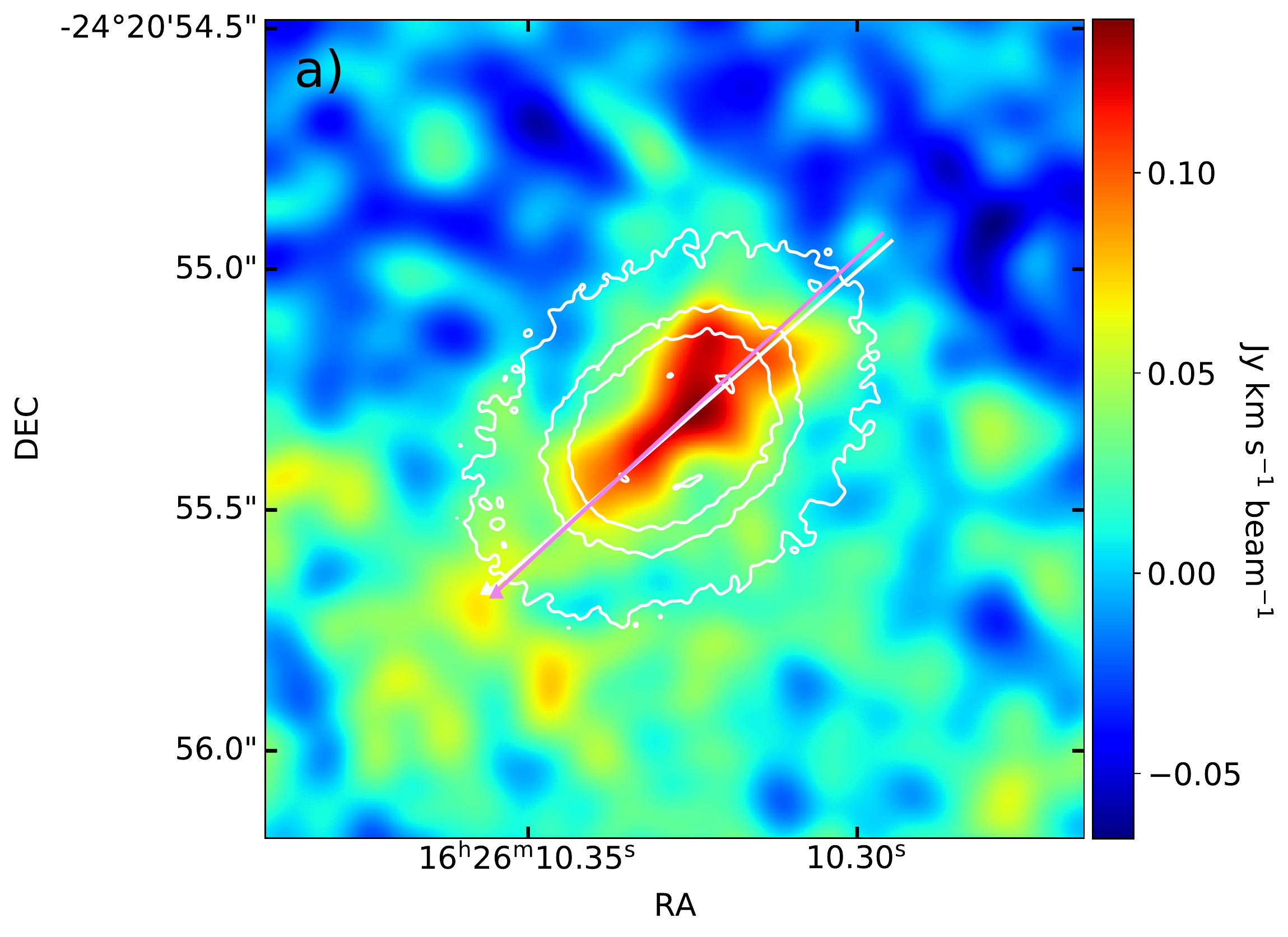} 
\includegraphics[width=0.435\textwidth]{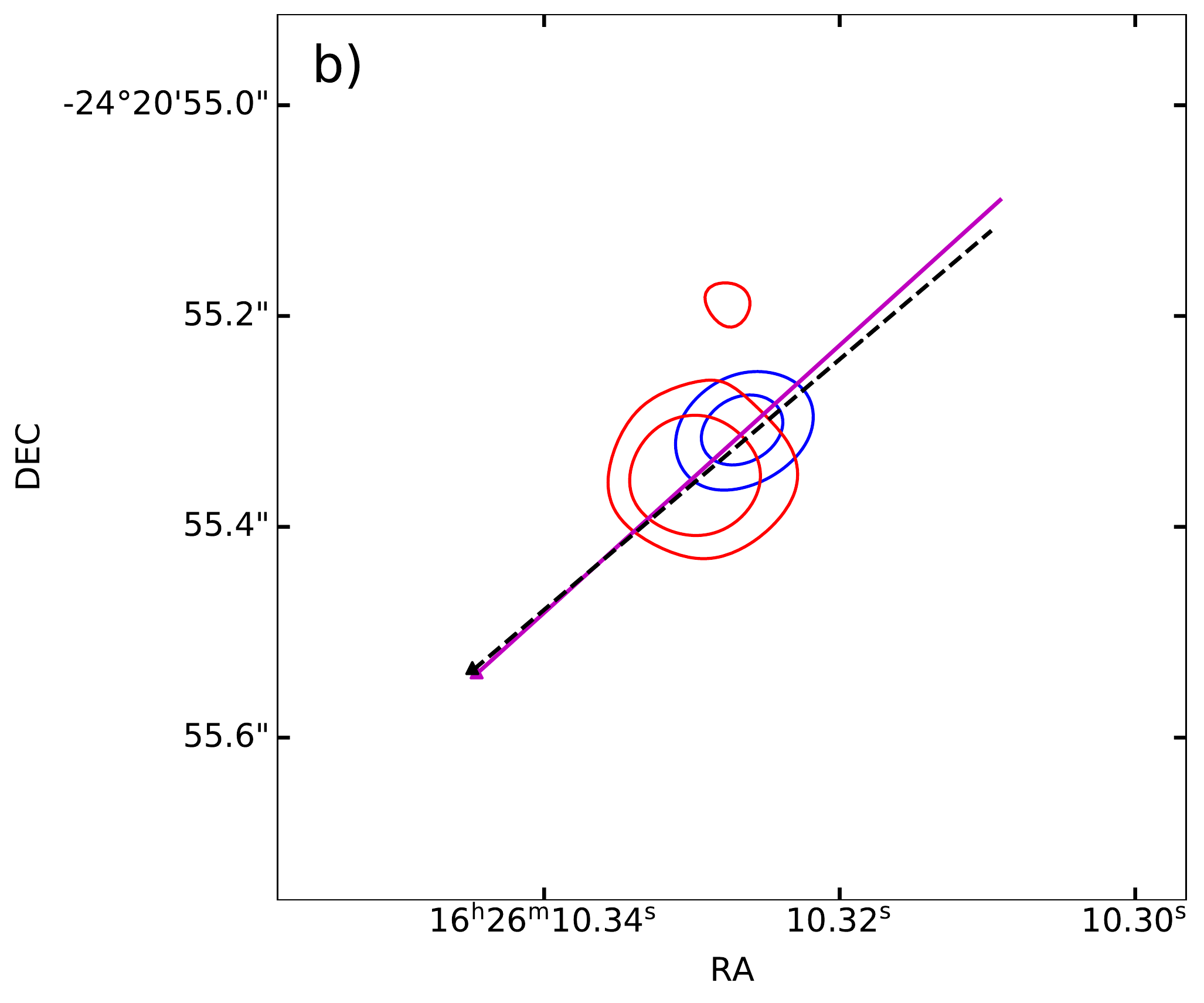}
\includegraphics[width=0.498\textwidth]{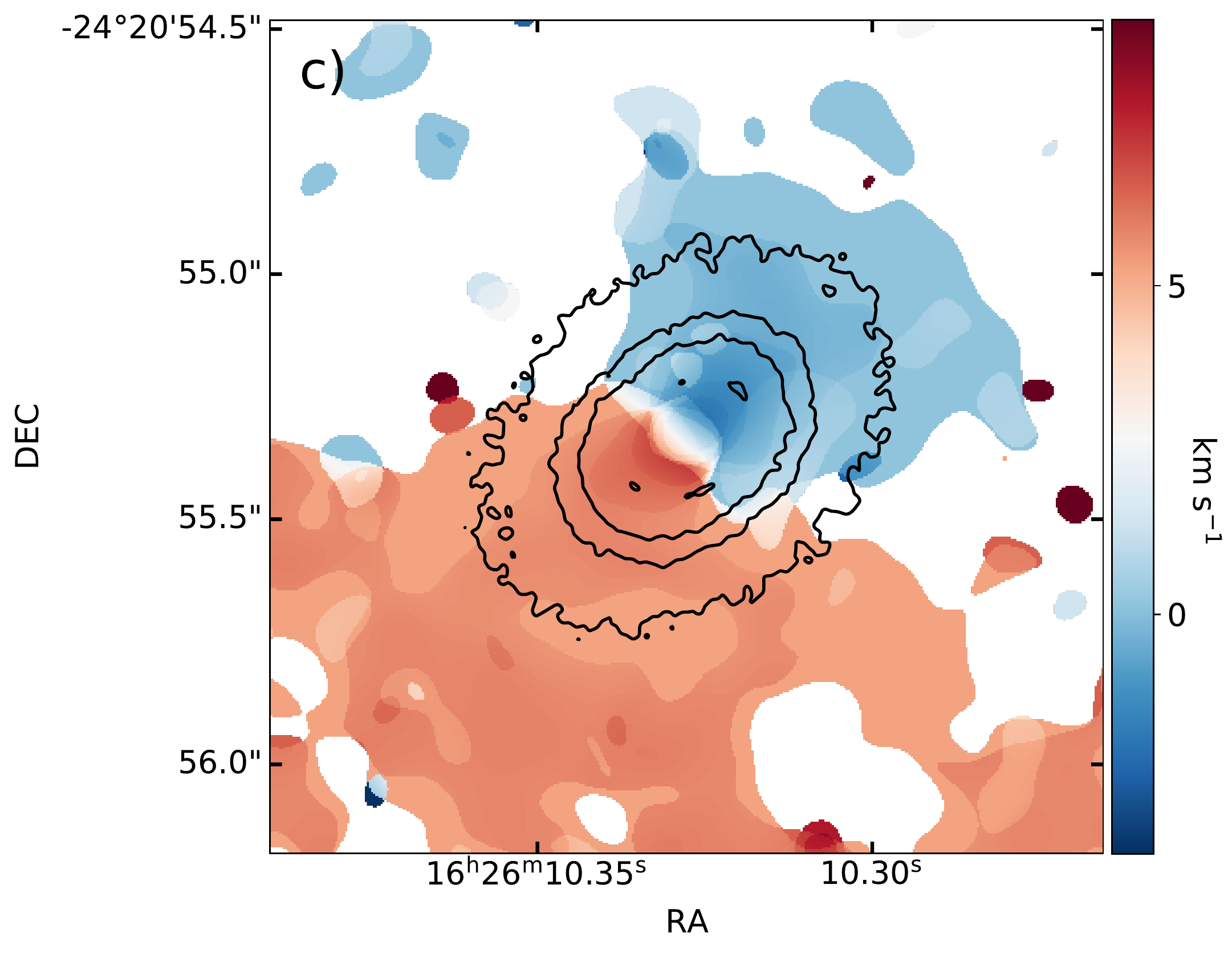}
\caption{(a) Moment 0 map of the $^{12}$CO line for ISO-Oph\,17, the integrated velocities range from -4.92 to 9.05~km~s$^{-1}$, contours of continuum emission at the 5$\sigma$ level are indicated in white. (b) Contours represent 6.35 and 10~mJy beam$^{-1}$. (c) Moment 1 map of $^{12}$CO emission for ISO-Oph\,17, the integrated velocities range from -4.92 to 9.05~km~s$^{-1}$. Annotations follow from Fig.\,\ref{fig:doarPA}.}
\label{fig:pa-iso-oph-17}
\end{figure*}

\begin{figure*}
\centering
    \includegraphics[width=15.7cm]{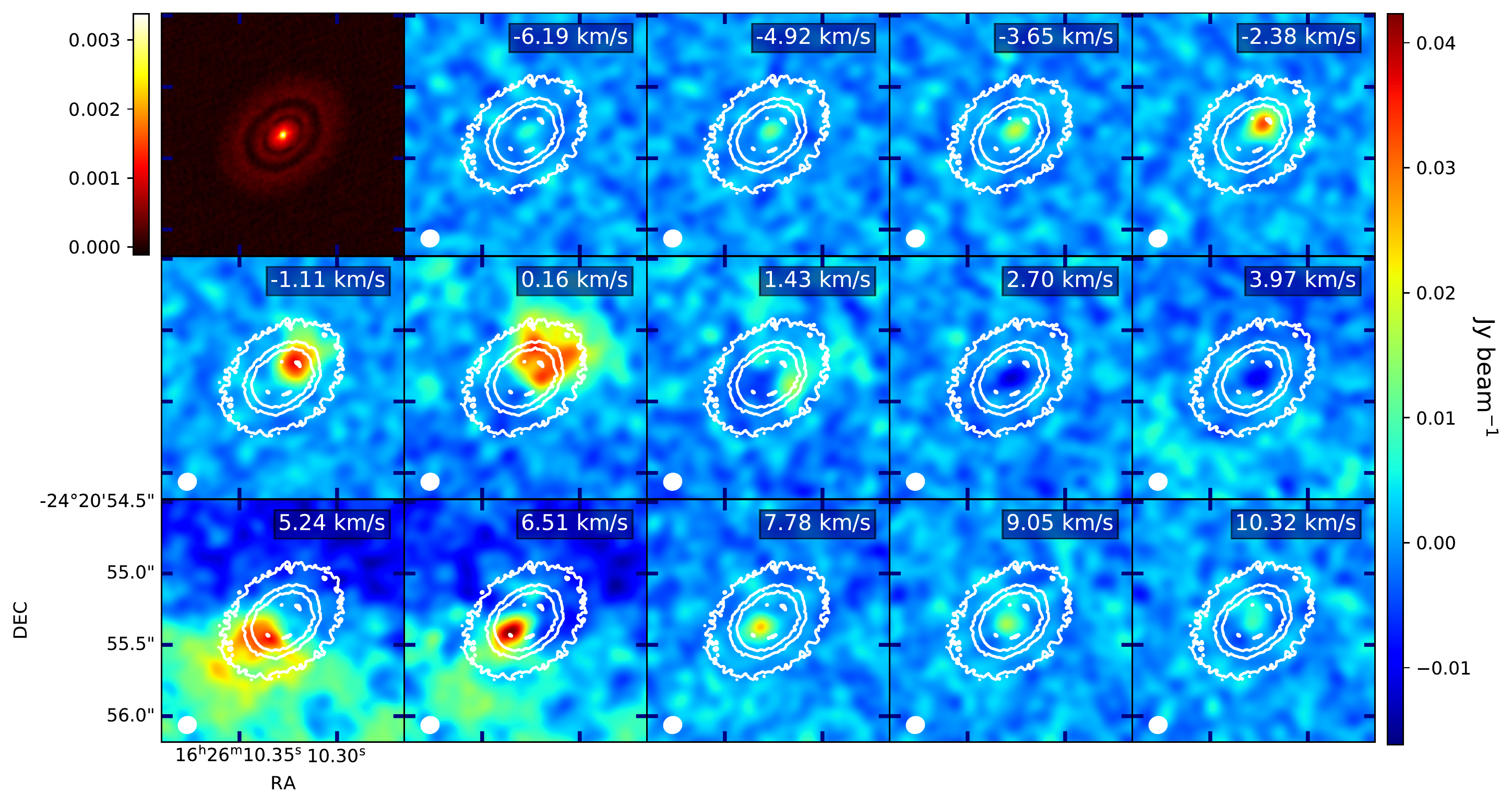}
    \caption{Observed channel maps of $\protect {}^{12}$CO(2-1) of ISO-Oph 17. Contours of 1.3 mm continuum emission at the 5$\sigma$ level are shown in white. Annotations follow from Fig.\,\ref{fig:my_labelDO}.}
    \label{fig:my_label17}
\end{figure*}

This source is also known as GSS 26m and is a Class II source.\\
Previous ALMA observations of the dust have resolved the structure of its disc, revealing concentric rings and gaps.\\
Our observations in Fig.\,\ref{fig:pa-iso-oph-17} and Fig.\,\ref{fig:my_label17} show $^{12}$CO(2-1) emission in the inner and outer parts of the disc. No significant difference was found between the P.A. measured for the gas and the P.A. of the dust (Fig.\,\ref{fig:pa-iso-oph-17}). We also identify contamination from the molecular cloud in all the channels. 

\subsubsection{SR\,24S}
\begin{figure*}
\includegraphics[width=0.498\textwidth]{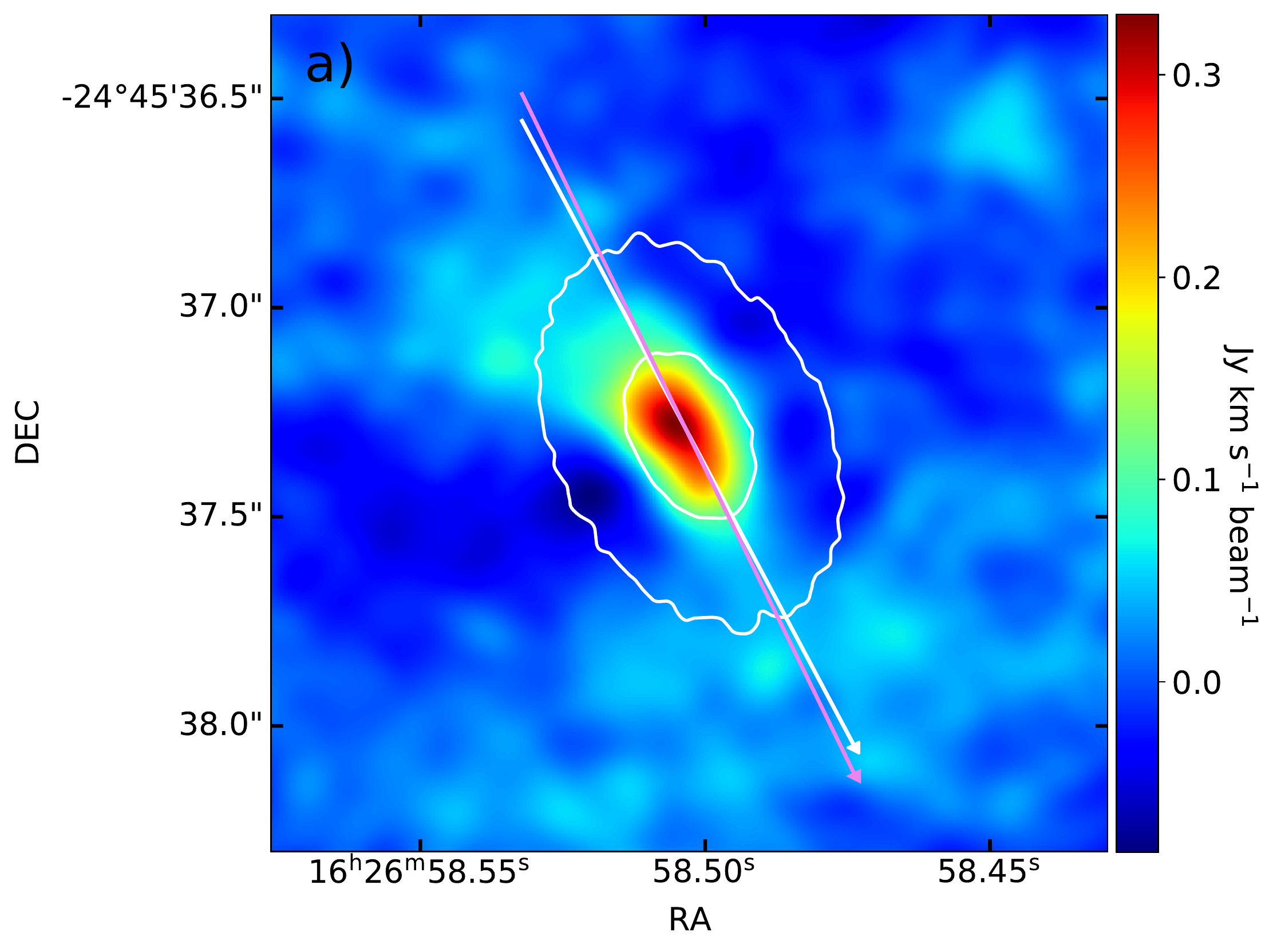} 
\includegraphics[width=0.448\textwidth]{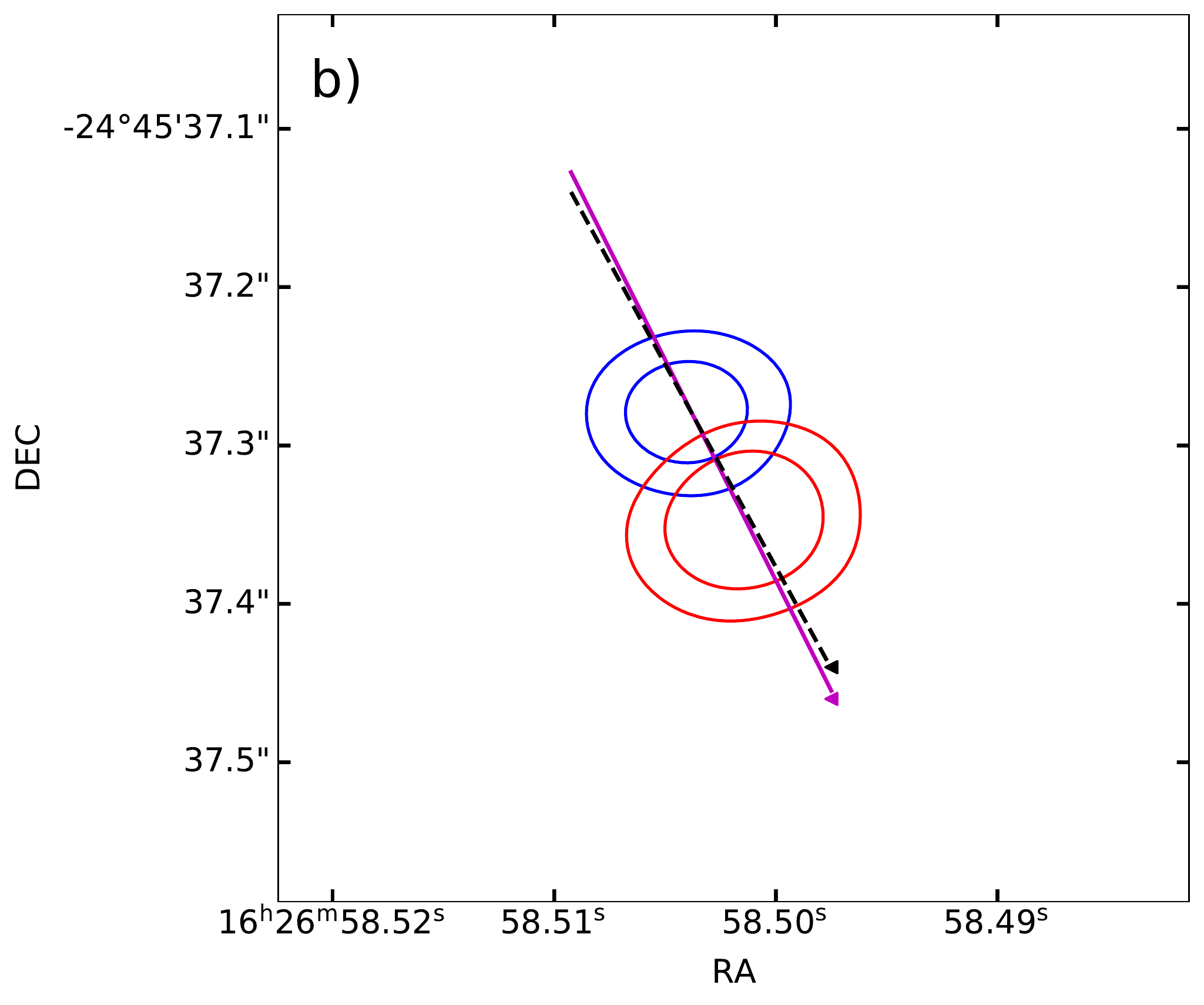}
\includegraphics[width=0.448\textwidth]{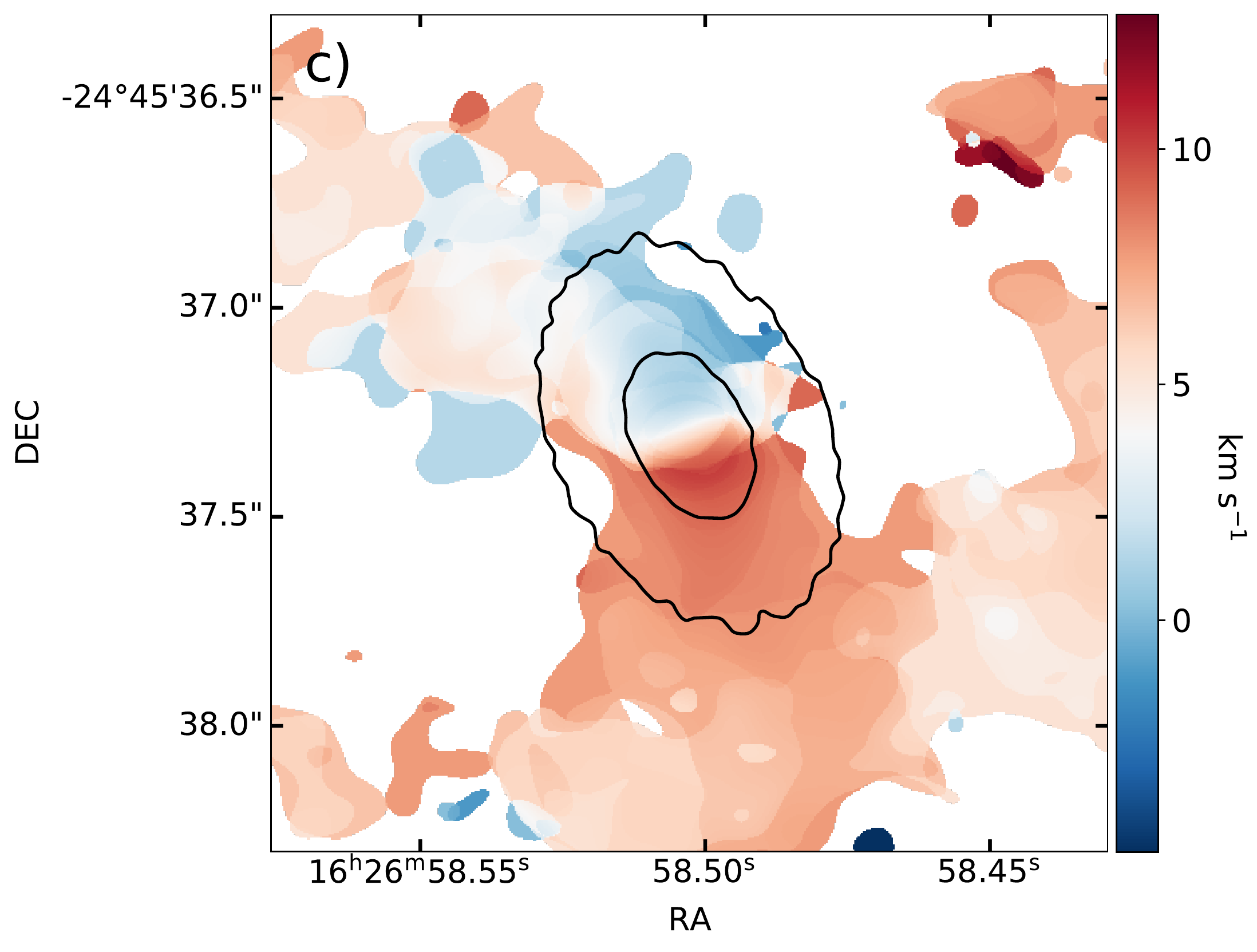}
\caption{(a) Moment 0 map of the $^{12}$CO line for SR\,24S, the integrated velocities range from -4.92 to 14.13~km~s$^{-1}$, contours of continuum emission at the 20$\sigma$ level are indicated in white. (b) Contours represent 8.6 and 11.6~mJy beam$^{-1}$. (c) Moment 1 map of $^{12}$CO emission for SR\,24S, the integrated velocities range from -4.92 to 14.13~km~s$^{-1}$, contours of continuum emission at the 20$\sigma$ level are indicated in black. Annotations follow from Fig.\,\ref{fig:doarPA}.}
\label{fig:srPA}
\end{figure*}

\begin{figure*}
\centering
    \includegraphics[width=15.7cm]{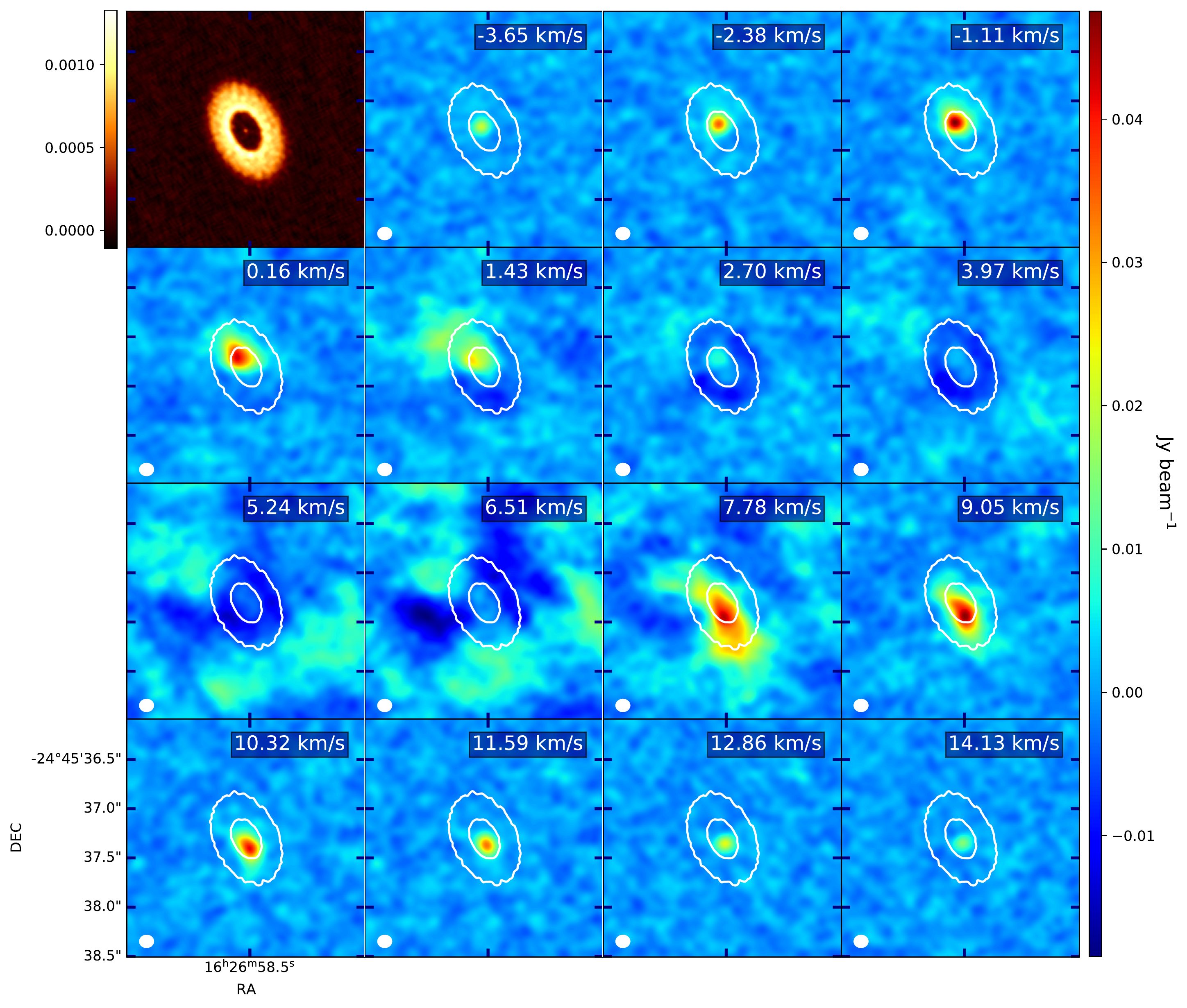}
    \caption{Observed channel maps of $\protect {}^{12}$CO(2-1) of SR\,24S. Contours of 1.3 mm continuum emission at the 20$\sigma$ level are shown in white. Annotations follow from Fig.\,\ref{fig:my_labelDO}.}
    \label{fig:my_labelSR}
\end{figure*}

This source is a single star part of a hierarchical triple system, where SR 24N is a binary system. Both circumstellar discs have been detected at 1.3 mm, and $^{12}$CO(2-1) observations have shown evidence of a bridge of gas joining both discs \citep{Fern_ndez_L_pez_2017}. NIR observations have revealed a bridge of infrared emission connecting SR\,24N and SR\,24S \citep{Mayama2010}.
Previous continuum observations of SR\,24S have shown a structure of ring \citep{Pinilla_2017, Cieza2021}. \citet{Pinilla_2017} detected $^{13}$CO and C$^{18}$O  (2$-$1) emission which peaks at the centre of the continuum cavity of SR\,24S. Besides, \citet{VanDerMarel2015} reported gas emission inside the dust cavity of SR\,24S from observations of the $^{12}$CO(6$-$5) line.\\
In our long-baseline observations, we detect gas in SR\,24S and SR 24N but not a bridge of gas connecting both discs because its size is longer than our maximum recoverable scale (MRS) $\sim$\,0.67 arcsec. Fig.\,\ref{fig:srPA} shows integrated intensity and weighted-velocity maps of $^{12}$CO emission for SR\,24S, along with a comparison of P.A. for the gas and dust. No significant difference was found between the P.A. measured for the gas in this work and that previously obtained for the dust. In SR\,24S, we identify $^{12}$CO(2$-$1) emission inside the inner gap of the disc and in the outer disc in the channel maps of Fig.\,\ref{fig:my_labelSR}. Moreover, we see the continuum in absorption in five channels between the velocities 1.43 and 6.51~km~s$^{-1}$, and those channels exhibit contamination from the cloud.

\subsection{Individual sources: discs without clear Keplerian features}

\subsubsection{WLY\,2-63}
\begin{figure*}
\includegraphics[width=0.49\textwidth]{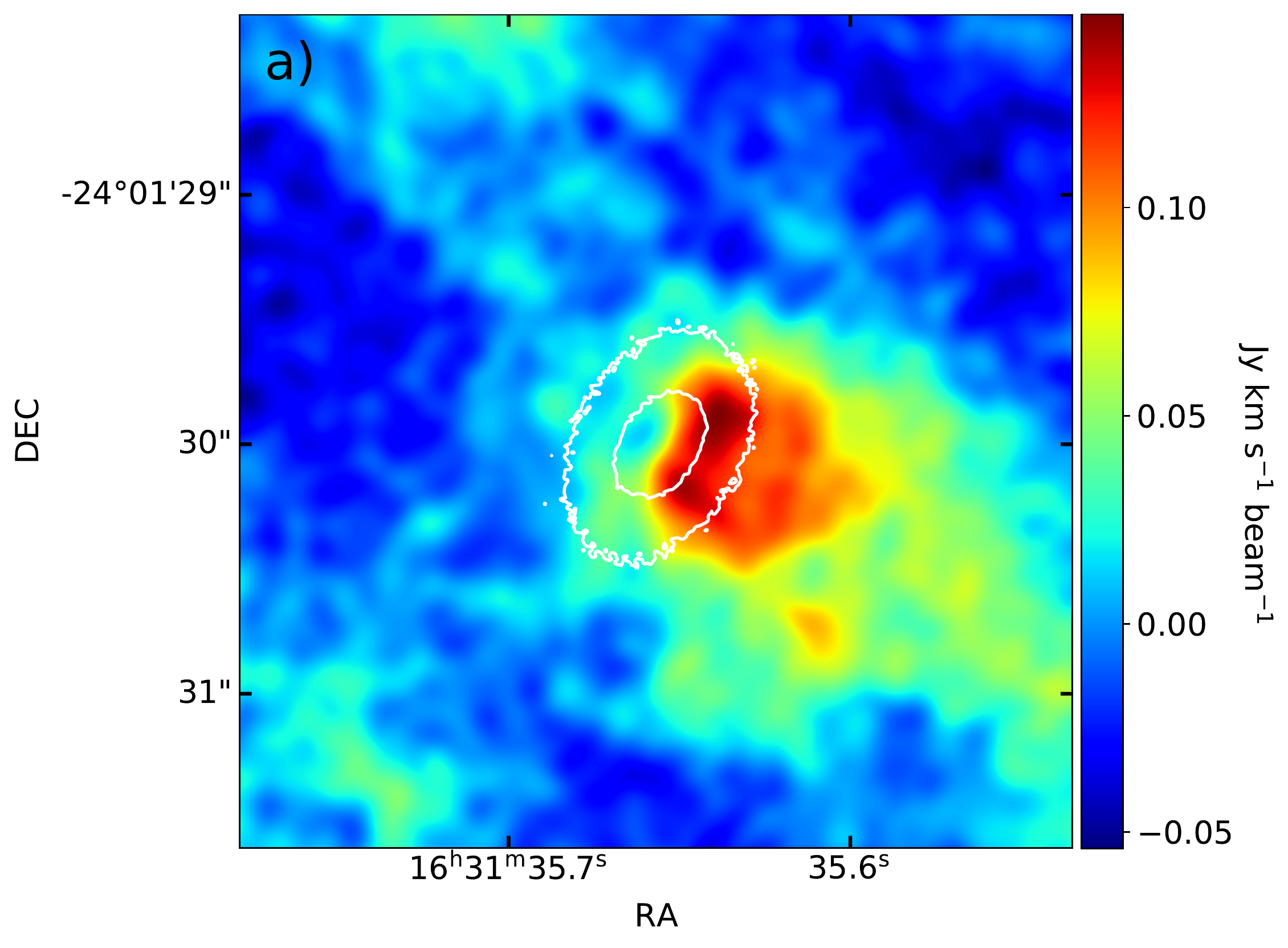}
\includegraphics[width=0.48\textwidth]{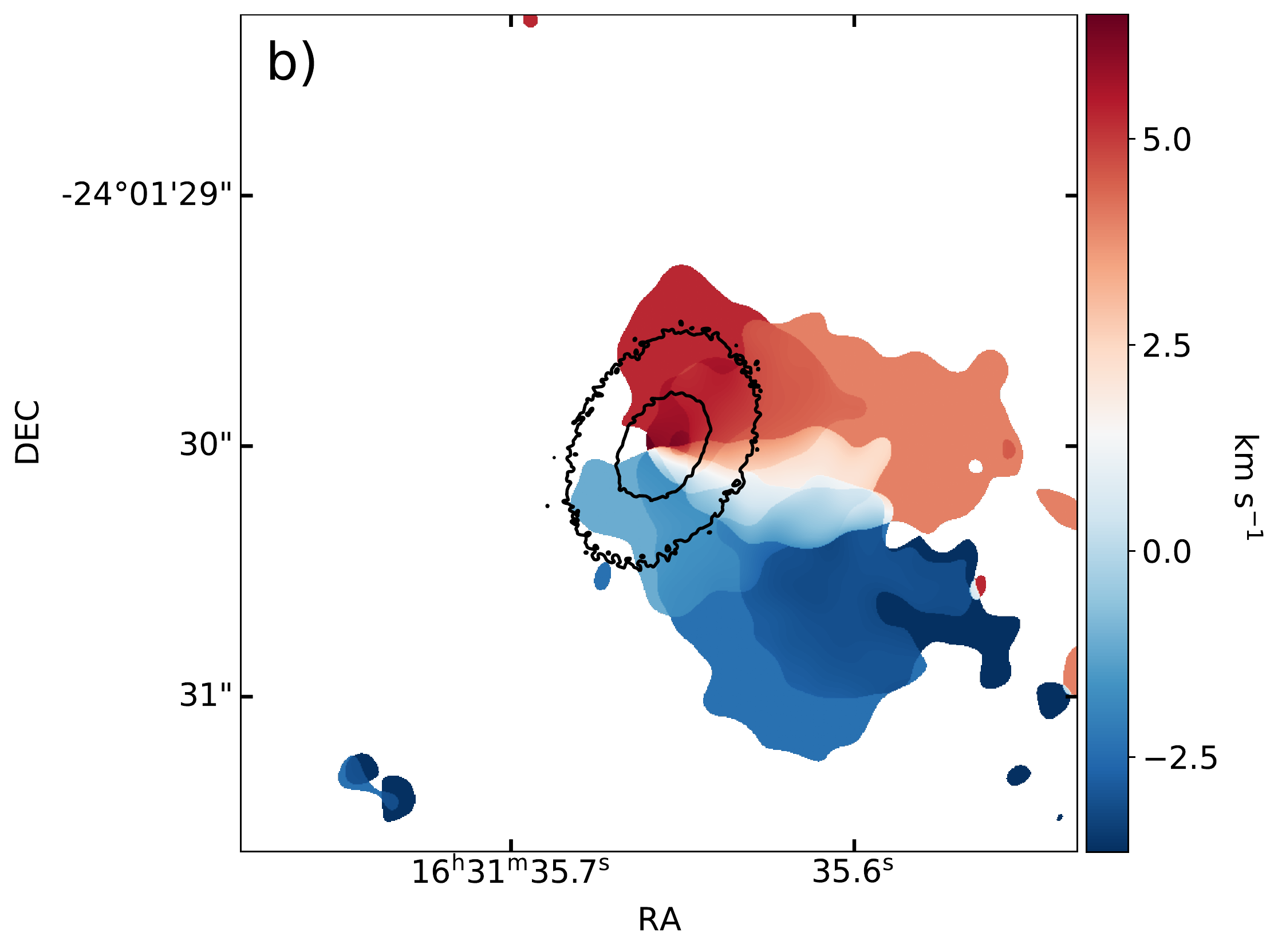}
\caption{(a) Moment 0 map of $^{12}$CO emission for WLY\,2-63, the integrated velocities range from -3.65 to -1.11~km~s$^{-1}$ and then from 3.97 to 7.78~km~s$^{-1}$, contours of 5$\sigma$ and 20$\sigma$ dust emission are shown in white. (b) Moment 1 map of $^{12}$CO emission for WLY\,2-63, the integrated velocities range from -3.65 to -1.11~km~s$^{-1}$ and then from 3.97 to 7.78~km~s$^{-1}$, contours of 5$\sigma$ and 20$\sigma$ dust emission are shown in black.}
\label{fig:mom0-maps}
\end{figure*}

\begin{figure*}
\centering
    \includegraphics[width=15.7cm]{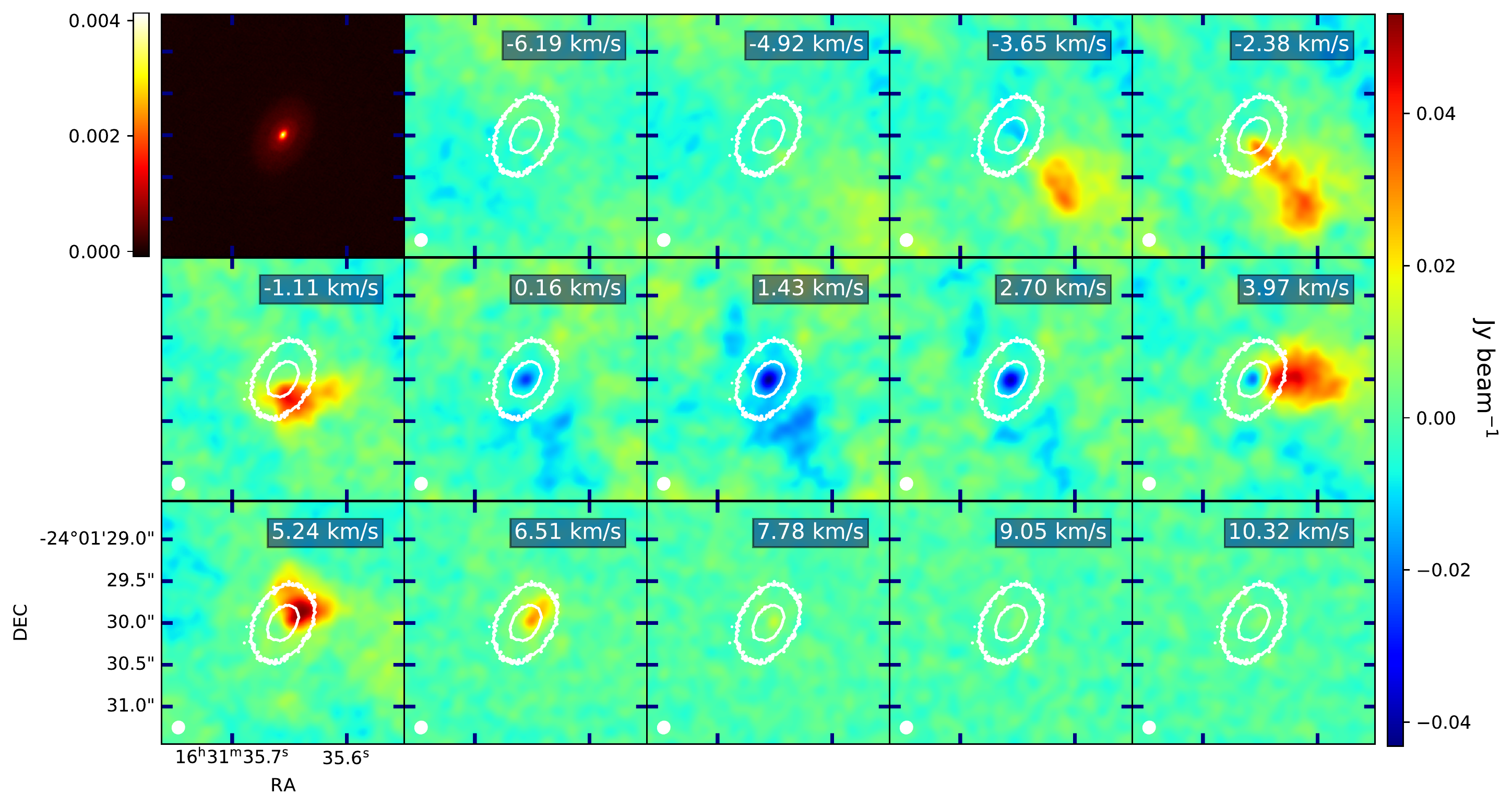}
    \caption{Observed channel maps of $\protect {}^{12}$CO(2-1) of WLY\,2-63. Contours of 1.3 mm continuum emission at the 5 and 20 $\sigma$ level are shown in white. Annotations follow from Fig.\,\ref{fig:my_labelDO}.}
    \label{fig:my_label263}
\end{figure*}

This source, also known as Oph Emb 17 and IRS 63, has a flat spectrum; it is the brightest target of the ODISEA long-baseline sample at 1.3 mm. The dust emission observations of \citet{SeguraCox2020} and \citet{Cieza2021} showed three concentric rings, two bright annular substructures, and two dark annular substructures. In addition, \citet{SeguraCox2020} noticed that the temperature of the outer inflection point is comparable to the condensation temperature of the CO.\\
In this work, the $^{12}$CO(2-1) emission reveals signatures that may be attributed to an outflow and could be confirmed with higher sensitivity observations at larger angular scales (González-Ruilova in prep.). We observe rotating gas with extended and asymmetric emission towards the southwest, showing its largest spatial extension in the channels labeled with the velocities -3~km~s$^{-1}$ and -2.38~km~s$^{-1}$ (moment maps in Fig.\,\ref{fig:mom0-maps} and channel maps in Fig.\,\ref{fig:my_label263}). We identify that part of the blue and red side of the Keplerian pattern of rotation appears absorbed, as we cannot clearly identify a counterpart for the emission in the observed channels.\\
We see the continuum in absorption in four channels, between the velocities 0.16~km~s$^{-1}$ and 3.97~km~s$^{-1}$ in the innermost region of the disc. This effect is due to the continuum subtraction process and the fact that interferometers do not sample homogeneous screens.

\subsubsection{ISO-Oph 37}
\begin{figure*}
\includegraphics[width=0.49\textwidth]{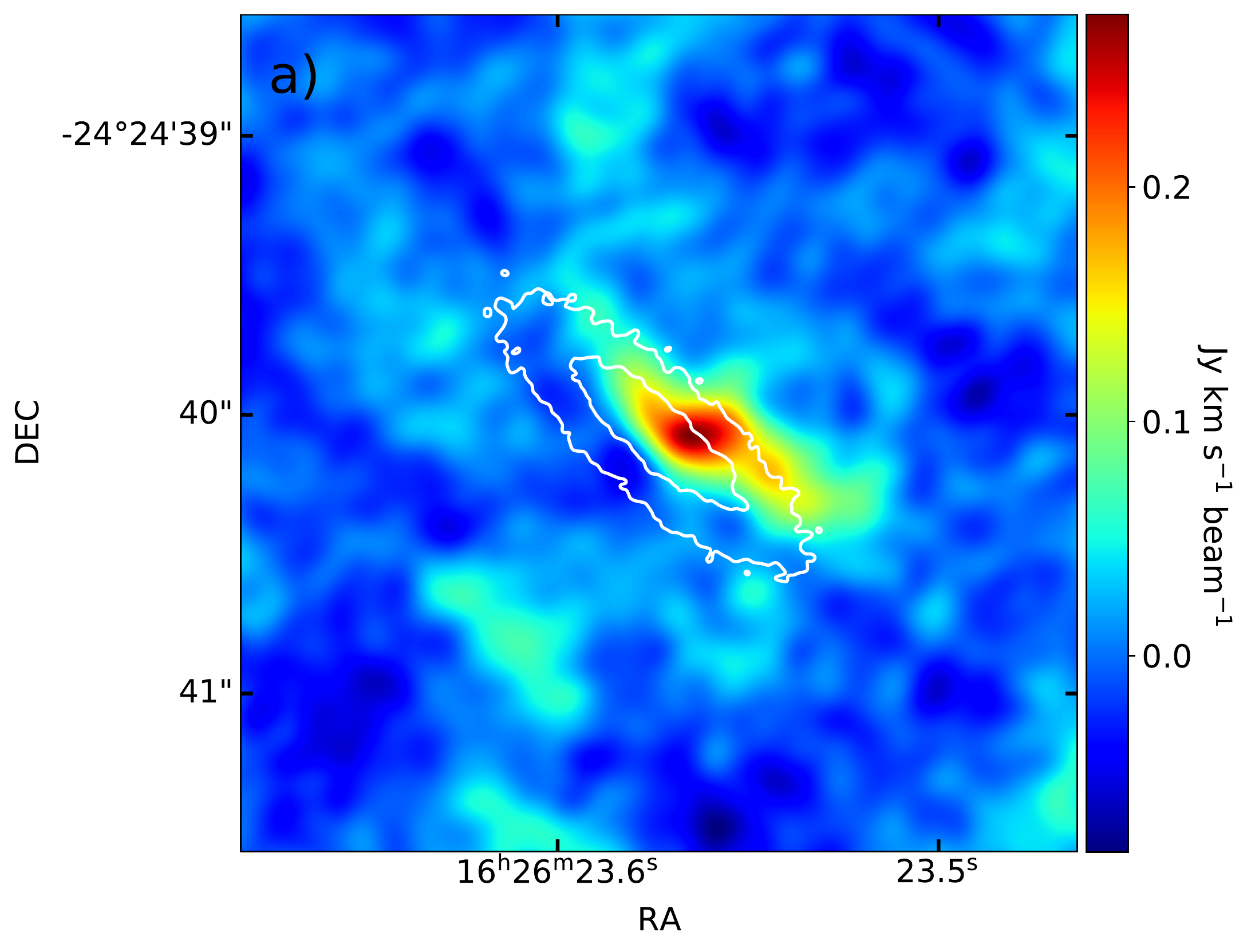}
\includegraphics[width=0.482\textwidth]{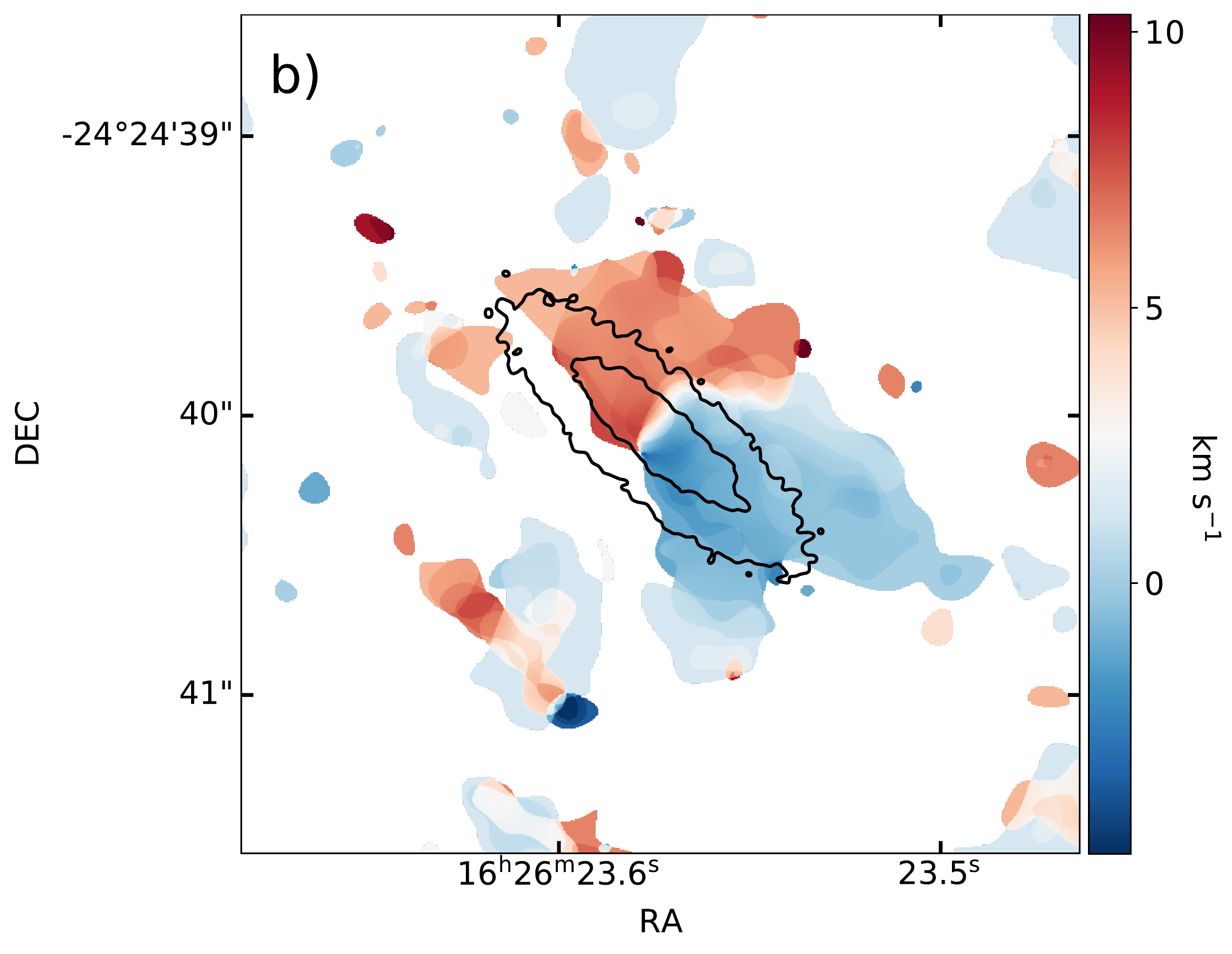}
\caption{(a) Moment 0 map of $^{12}$CO emission for ISO-Oph\,37, the integrated velocities range from -6.19 to 10.32~km~s$^{-1}$, contours of 5$\sigma$ and 20$\sigma$ dust emission are shown in white. (b) Moment 1 map of $^{12}$CO emission for ISO-Oph\,37, the integrated velocities range from -6.19 to 10.32~km~s$^{-1}$, contours of 5$\sigma$ and 20$\sigma$ dust emission are shown in black.}
\label{fig:mom-maps-iso-oph-37}
\end{figure*}

\begin{figure*}
\centering
    \includegraphics[width=15.7cm]{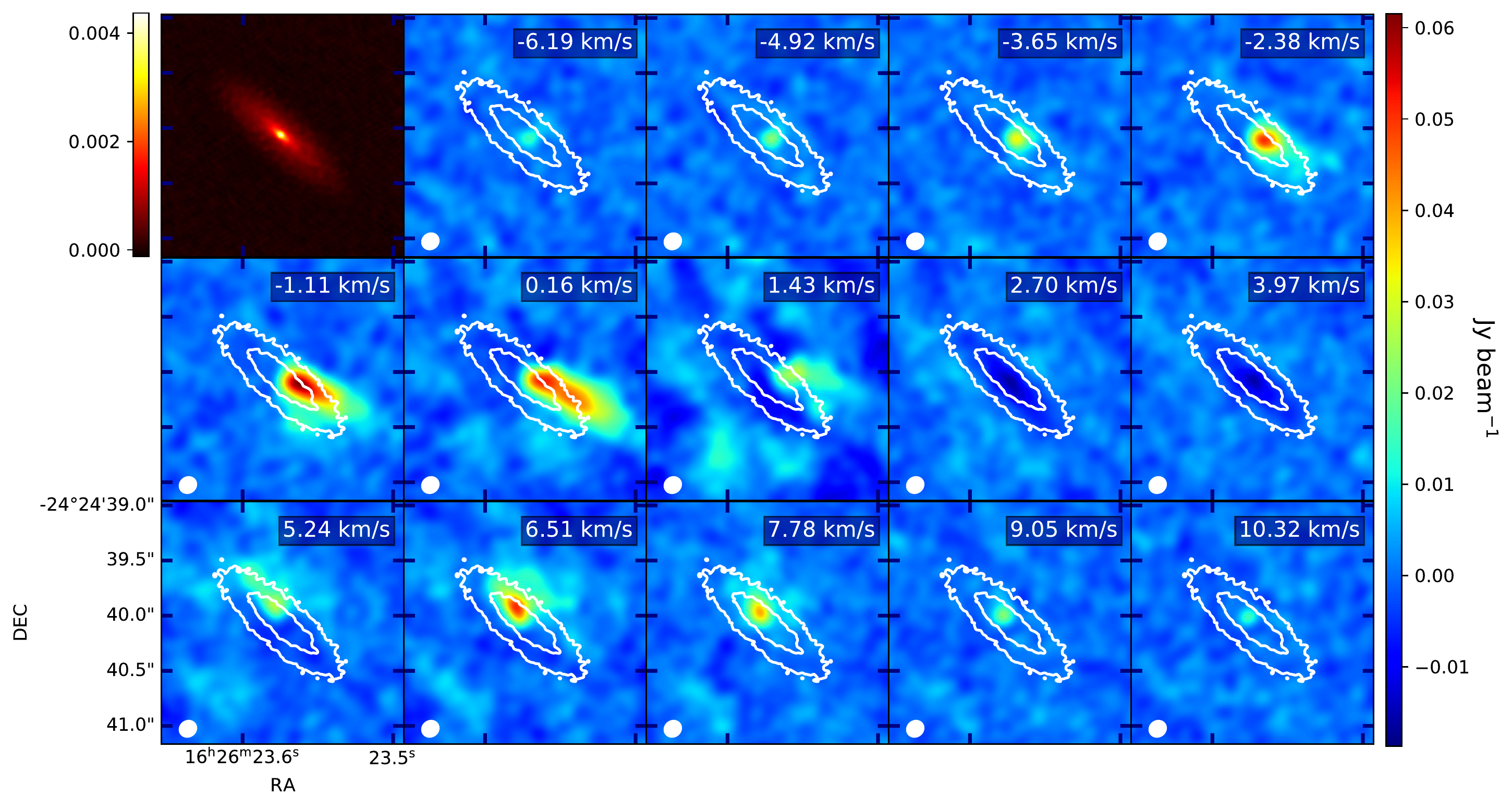}
    \caption{Observed channel maps of $\protect {}^{12}$CO(2-1) of ISO-Oph 37. Contours of 1.3 mm continuum emission at the 5 and 20$\sigma$ level are shown in white.  Annotations follow from Fig.\,\ref{fig:my_labelDO}.}
    \label{fig:my_label37}
\end{figure*}

This source is also known as GY 21 and has a flat spectrum SED. The ODISEA observations \citep[][]{Cieza2021} at 1.3 mm do not reveal any substructure on the radial axis, only an inflection point at 31 au.\\ Our observations in Fig.\,\ref{fig:mom-maps-iso-oph-37} and Fig.\,\ref{fig:my_label37} display gas emission in the inner and outer parts of the disc, and display asymmetric morphology of the gas in comparison with the position of the dust. In the channel maps of Fig.\,\ref{fig:my_label37} we do not observe a counterpart for the emission located towards the south-west direction at the velocities -1.11~km~s$^{-1}$ and 0.16~km~s$^{-1}$ (direction north-east). The offset from the disc midplane is due to the projection onto the plane of the sky of the altitude of the CO emitting layer above the disc midplane. Further, in the channel maps, we identify emission most likely coming from the cloud material (seen in the channel labeled with the velocity 1.43~km~s$^{-1}$). Continuum emission is also seen in absorption in the channels at 2.7 and 3.97~km~s$^{-1}$.

\subsubsection{ISO-Oph 54}
\begin{figure*}
\includegraphics[width=0.49\textwidth]{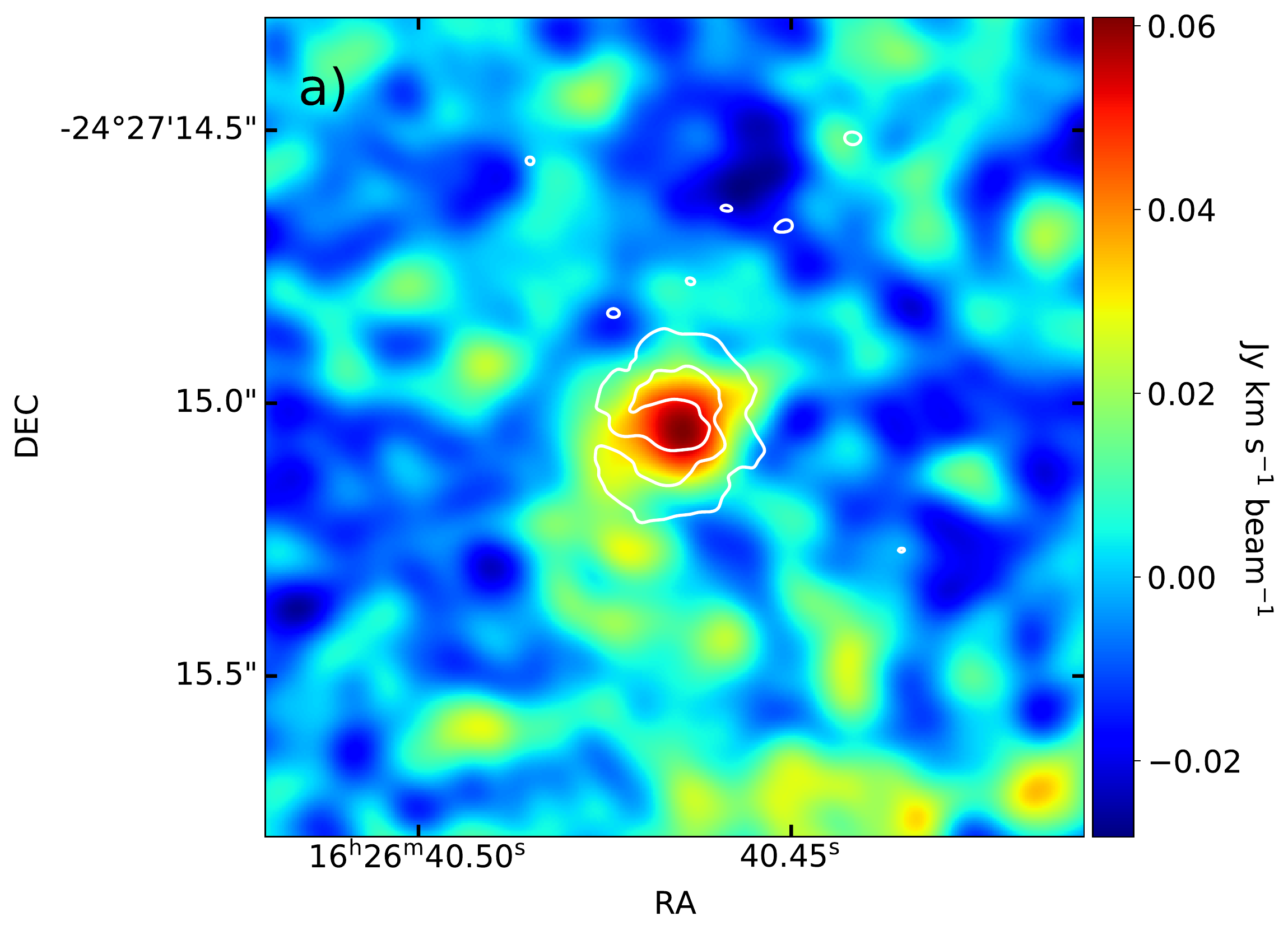}
\includegraphics[width=0.467\textwidth]{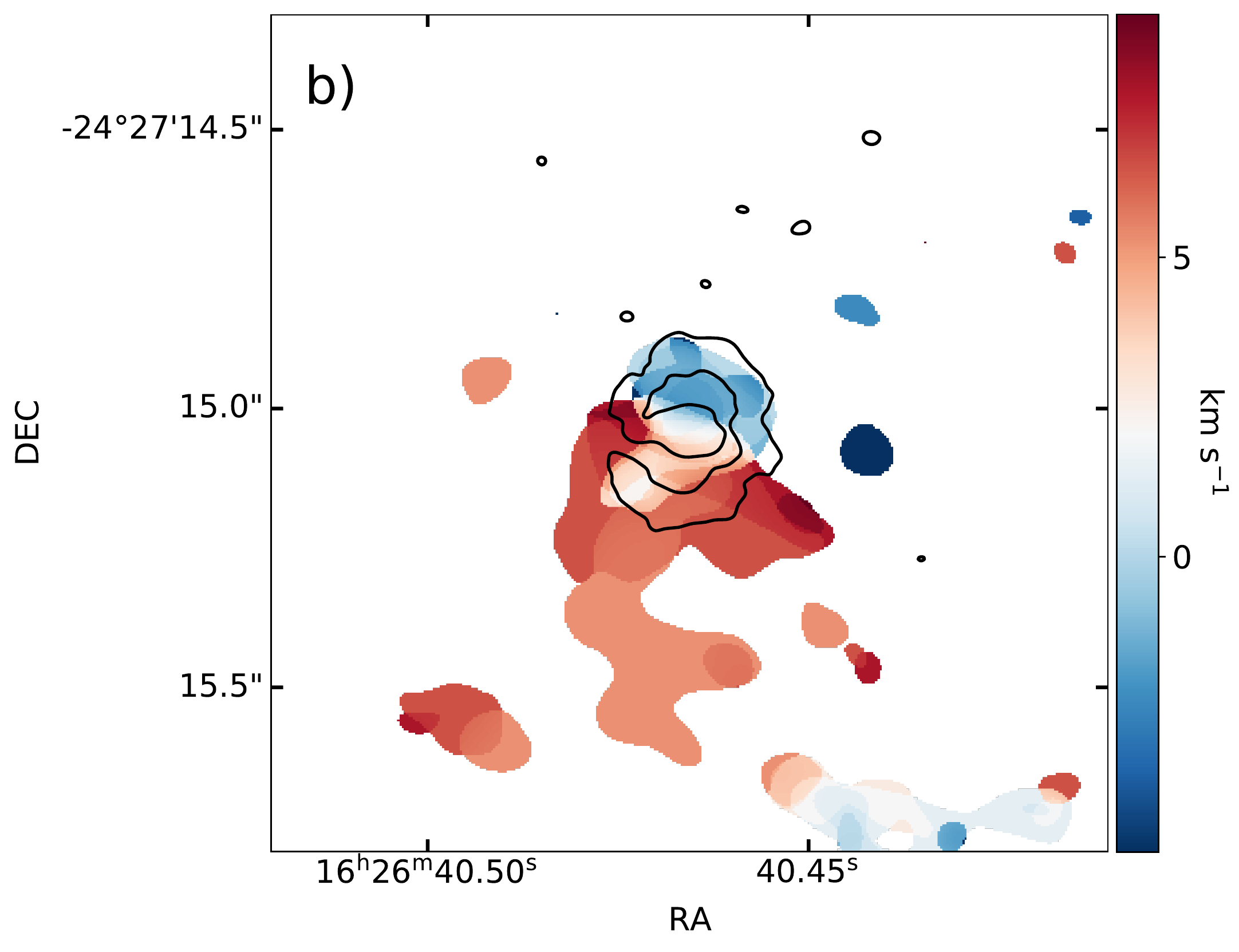}
\caption{(a) Moment 0 map of $^{12}$CO emission for ISO-Oph\,54, the integrated velocities range from -4.92 to 9.05~km~s$^{-1}$, contours of continuum emission at the 9$\sigma$ level are indicated in white. (b) Moment 1 map of $^{12}$CO emission for ISO-Oph\,54, the integrated velocities range from -4.92 to 9.05~km~s$^{-1}$, contours of continuum emission at the 9$\sigma$ level are indicated in black.}
\label{fig:mom-maps-iso-oph-54}
\end{figure*}

\begin{figure*}
\centering
    \includegraphics[width=15.7cm]{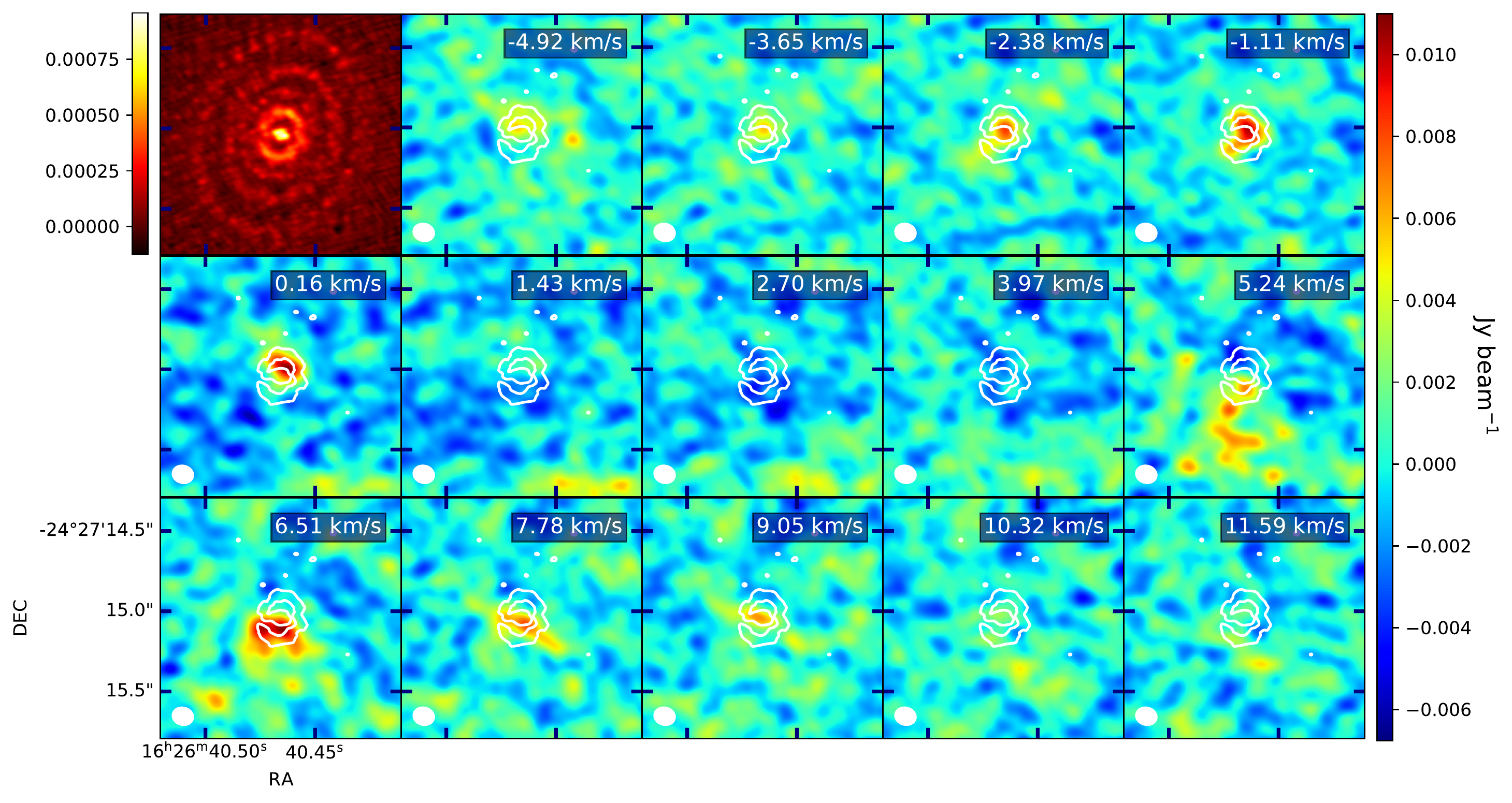}
    \caption{Observed channel maps of $\protect {}^{12}$CO(2-1) of ISO-Oph 54. Contours of 1.3 mm continuum emission at the 9$\sigma$ level are shown in white. Annotations follow from Fig.\,\ref{fig:my_labelDO}.}
    \label{fig:my_label54}
\end{figure*}

This source also known as Oph Emb 22 and GY 91, has a Class I SED. Previous ALMA observations of the continuum have resolved several gaps on its disc and an inner dust cavity \citep{Sheehan2018,Cieza2021}. Our long-baseline observations in Fig.\,\ref{fig:mom-maps-iso-oph-54} and Fig.\,\ref{fig:my_label54} display gas emission in the innermost part of the disc only. Besides, we notice emission from the cloud material in all the channels, suggesting that this disc is deeply embedded. 

\subsubsection{ISO-Oph 196}
\begin{figure*}
\includegraphics[width=0.49\textwidth]{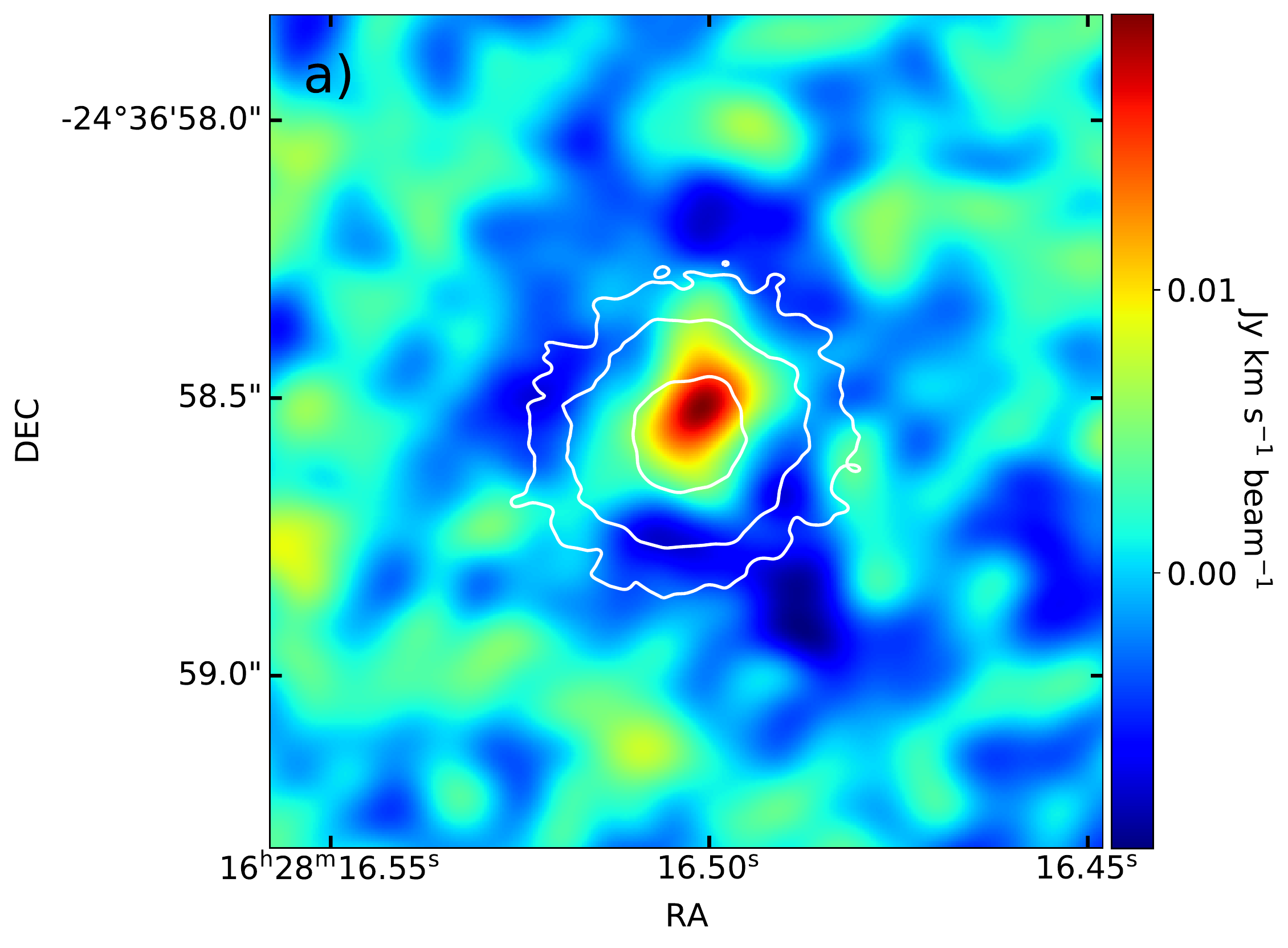}
\includegraphics[width=0.473\textwidth]{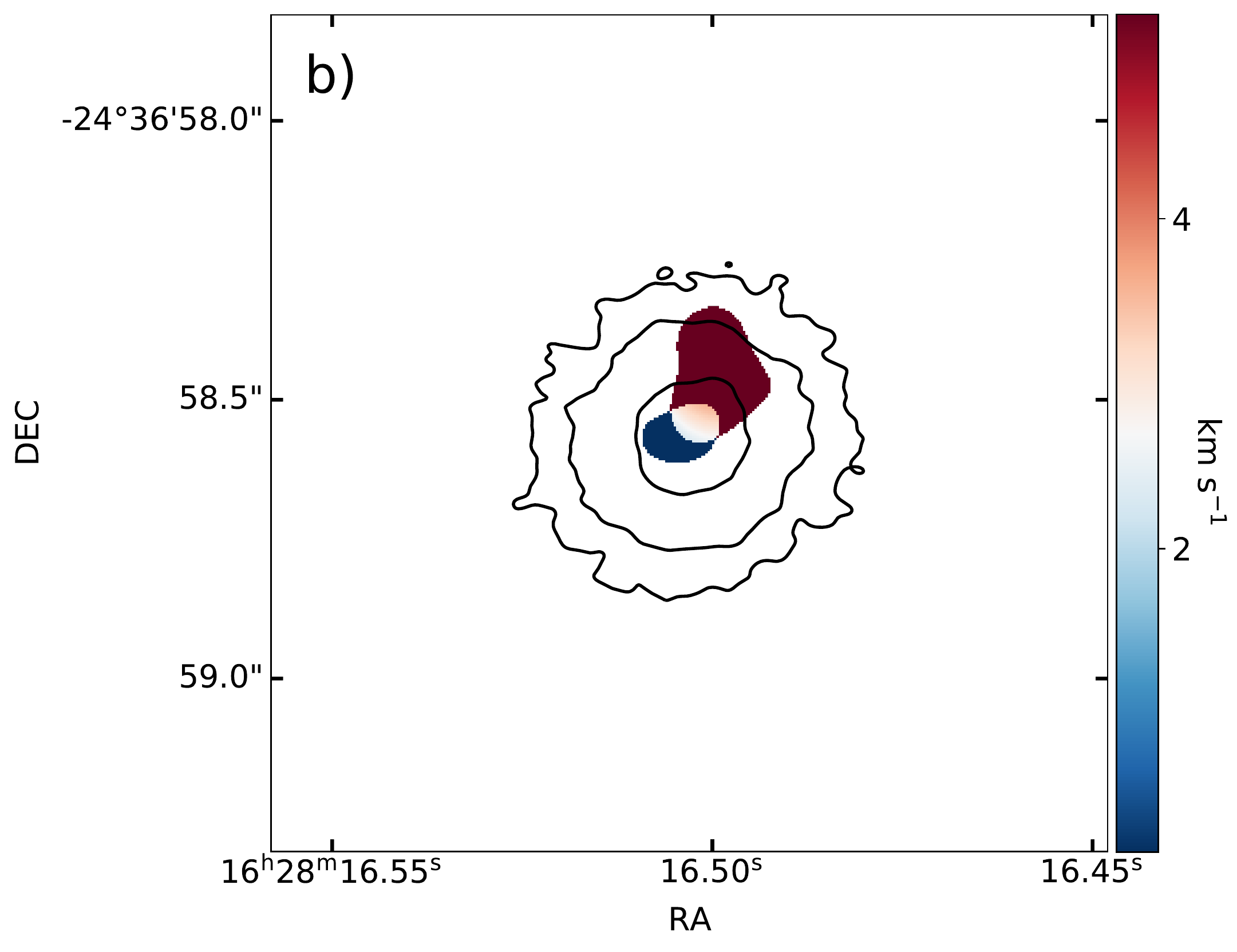}
\caption{(a) Moment 0 map of $^{12}$CO emission for ISO-Oph\,196, the integrated velocity range includes channels at velocities 0.16 and 5.24~km~s$^{-1}$, contours of continuum emission at the 9$\sigma$ level are indicated in white. (b) Moment 1 map of $^{12}$CO emission for ISO-Oph\,196, the integrated velocity range includes channels at velocities 0.16 and 5.24~km~s$^{-1}$, contours of continuum emission at the 9$\sigma$ level are indicated in black.}
\label{fig:mom-maps-iso-oph-196}
\end{figure*}

\begin{figure*}
\centering
    \includegraphics[width=15.7cm]{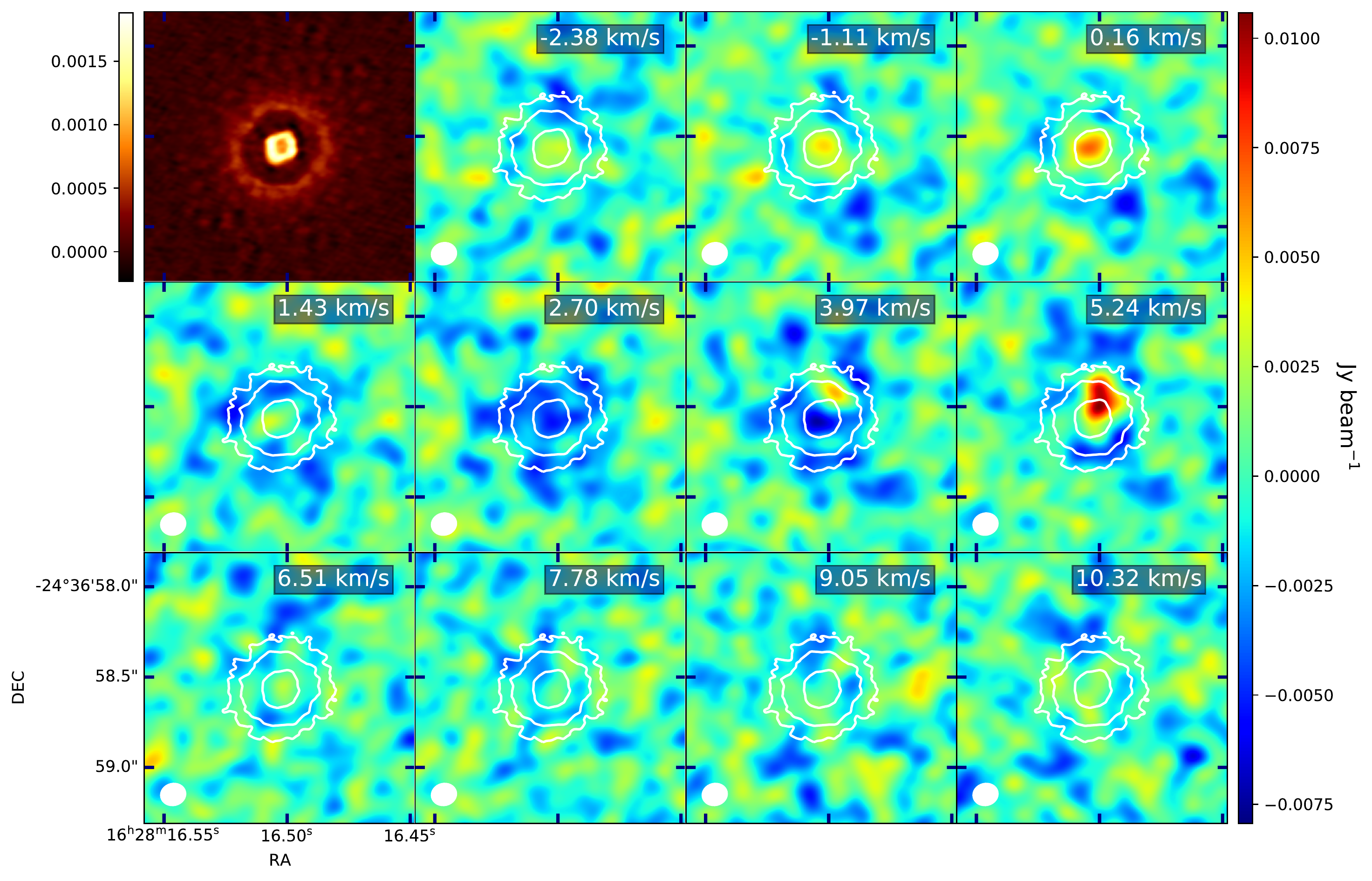}
    \caption{Observed channel maps of $\protect {}^{12}$CO(2-1) of ISO-Oph 196. Contours of 1.3 mm continuum emission at the 9$\sigma$ level are shown in white.  Annotations follow from Fig.\,\ref{fig:my_labelDO}.}
    \label{fig:my_label196}
\end{figure*}

This source is also known as WSB 60 and WLY\,1-58 and has a Class II SED. The dust observations of \citet{Francis2020} and \citet{Cieza2021} revealed an inner disc surrounded by a gap and a ring. Our long-baseline observations in Fig.\,\ref{fig:mom-maps-iso-oph-196} and Fig.\,\ref{fig:my_label196} show compact emission that is compatible with the position of the disc in two channels only, at 0.16~km~s$^{-1}$, and 5.24~km~s$^{-1}$. In the second channel (Fig.\,\ref{fig:my_label196}) the emission coincides with the position of the outer gap. We highlight the compact nature of the signal inside the gap, and without a symmetric counterpart about the disc major axis, as would be expected by diffuse gas inside the gap and in Keplerian rotation. Besides, we see the continuum in absorption in two channels, at the velocities 2.7~km~s$^{-1}$ and 3.97~km~s$^{-1}$.

\subsubsection{ISO-Oph 2}
\begin{figure*}
\includegraphics[width=0.49\textwidth]{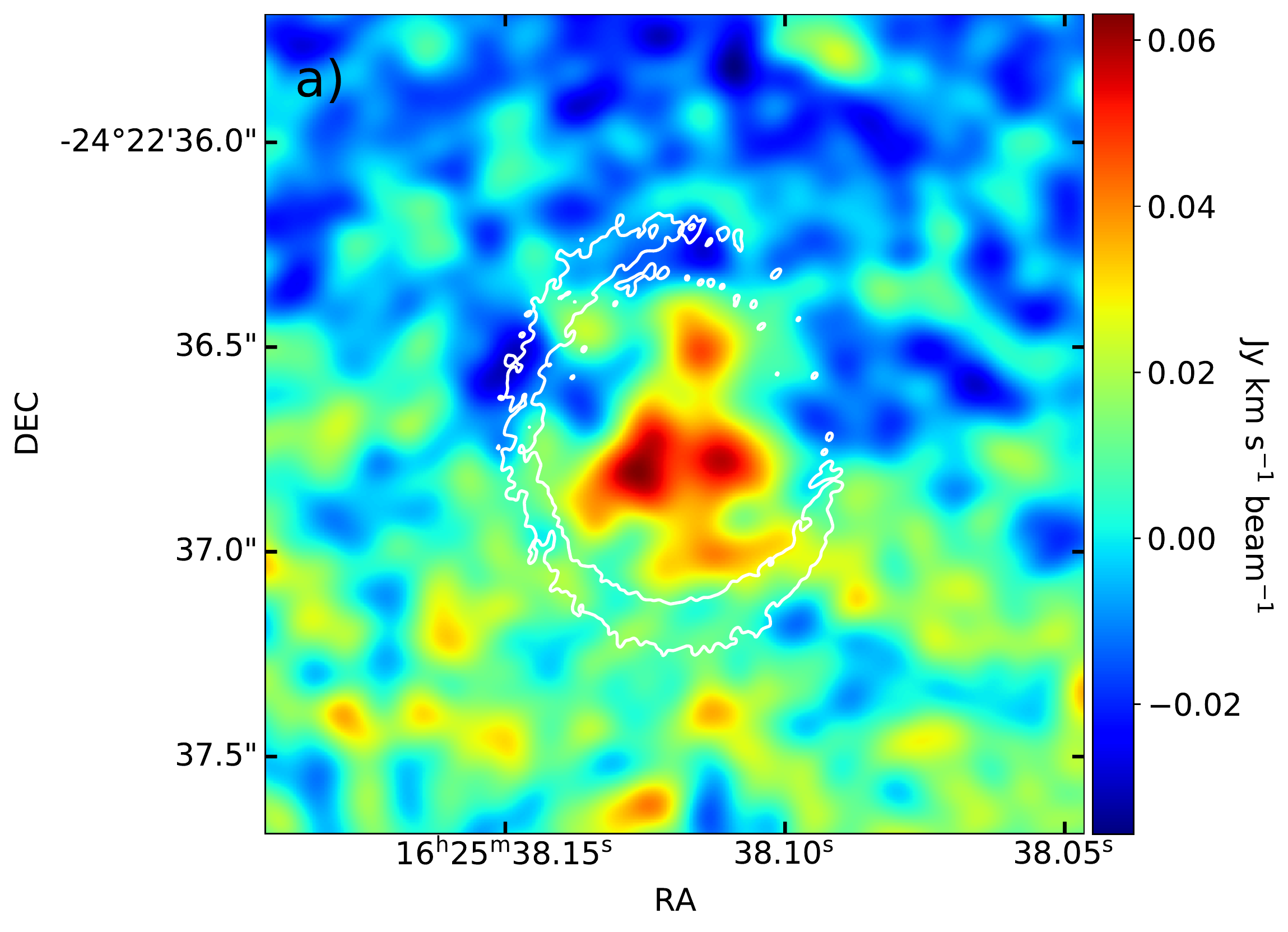}
\includegraphics[width=0.473\textwidth]{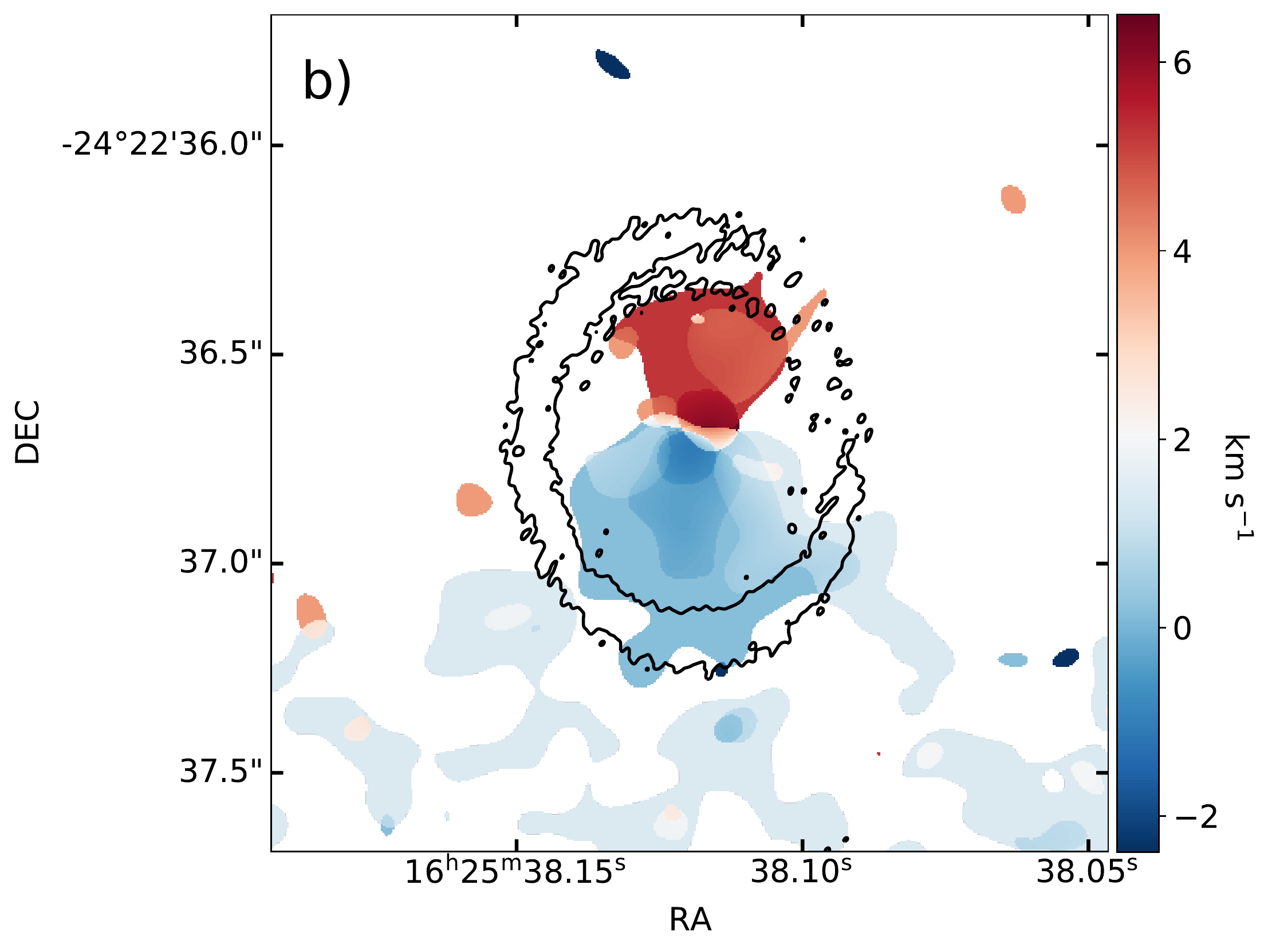}
\caption{(a) Moment 0 map of $^{12}$CO emission for ISO-Oph\,2, the integrated velocities range from -2.38 to 6.51~km~s$^{-1}$, contours of continuum emission at the 4$\sigma$ level are indicated in white. (b) Moment 1 map of $^{12}$CO emission for ISO-Oph\,2, the integrated velocities range from -2.38 to 6.51~km~s$^{-1}$, contours of continuum emission at the 4$\sigma$ level are indicated in black.}
\label{fig:mom-maps-iso-oph-2}
\end{figure*}

\begin{figure*}
\centering
    \includegraphics[width=15.7cm]{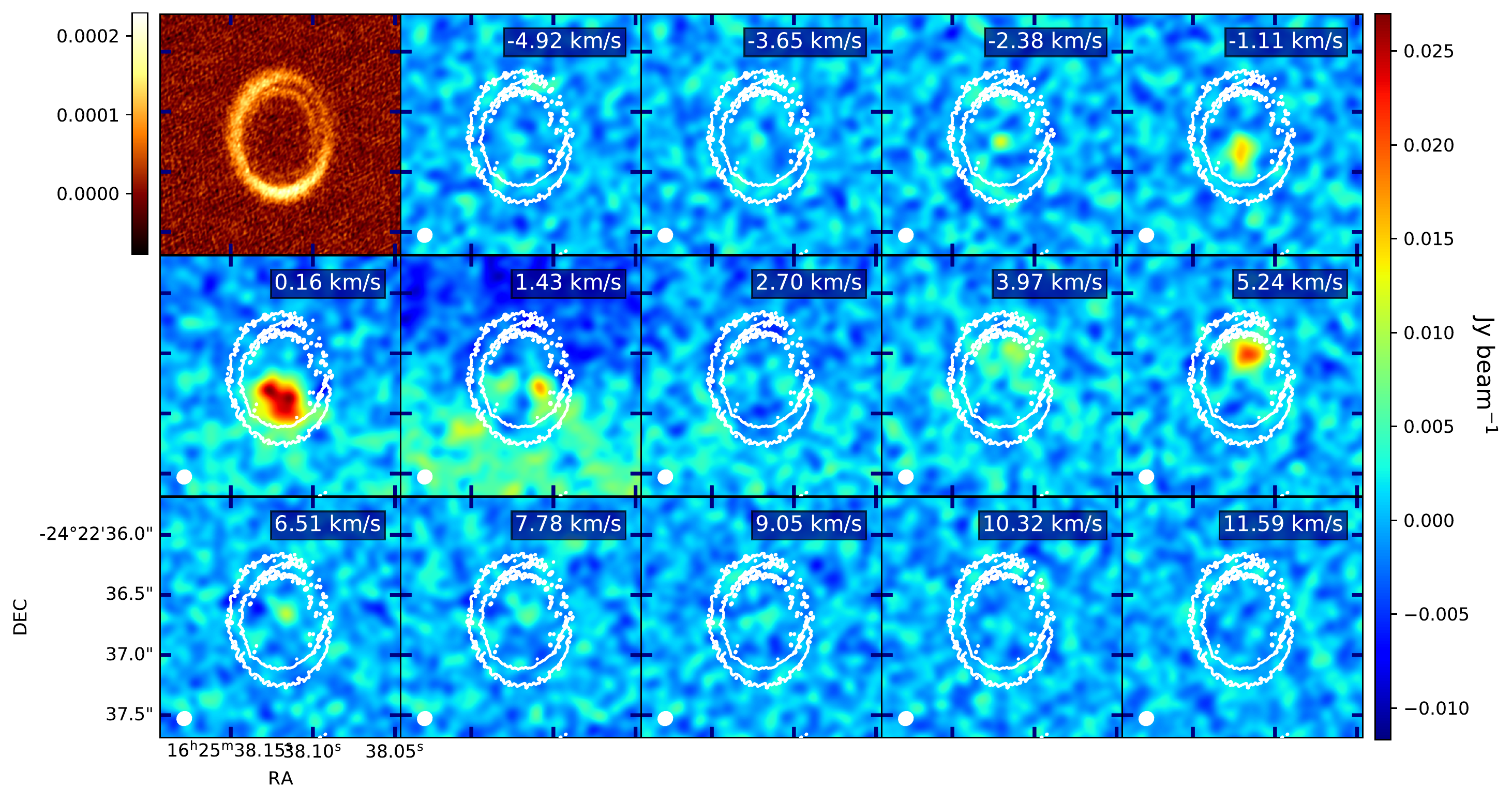}
    \caption{Observed channel maps of $\protect {}^{12}$CO(2-1) of ISO-Oph 2. Contours of 1.3 mm continuum emission at the 4$\sigma$ level are shown in white.  Annotations follow from Fig.\,\ref{fig:my_labelDO}.}
    \label{fig:my_label2}
\end{figure*}

ISO-Oph 2 is a binary system and has a Class II SED. Previous dust observations have resolved the primary disc displaying two nonaxisymmetric rings, and an inner cavity in the secondary disc. At the same time, $^{12}$CO(2-1) observations have revealed a possible bridge of gas connecting both discs. Our moment maps in Fig.\,\ref{fig:mom-maps-iso-oph-2} and our image in Fig.\,\ref{fig:my_label2} display the primary disc only. For a detailed analysis of this system see \citet{Gonzalez-Ruilova2020}.

\subsection{General discussion}
\label{sec:general-discussion}
Because the radio continuum is usually easier to detect than the lines, line surveys of discs are less common than for the dust component \citep[][]{Andrews2020, Bergin2017}; consequently, there have been considerably fewer studies of disc gas content than dust studies.\\
However, in recent years, there has been an increasing number of reports on gas properties such as total disc gas mass \citep[e.g.][]{Williams2014, Long2017}, gas density distributions \citep[e.g.][]{Fedele2017, Zhang2021}, temperature structures \citep[e.g.][]{Schwarz2016, Pinte2018}, abundances and distribution of molecules \citep[e.g.][]{Oberg2021, Guzman2021}, and kinematics \citep[e.g.][]{Perez2015, Teague2018, Pinte2019}.\\
There have also been significant advances in studies of the gas content of large disc populations in different star-forming regions, such as Lupus \citep[][]{Ansdell2016,Ansdell2018}, Taurus \citep[][]{Kurtovic2021,Rota2022}, Chamaleon II \citep[][]{Villenave2021}, and the Upper Scorpius OB association \citep[][]{Barenfeld2017}. This study focuses on the ODISEA long-baseline sample, which includes ten discs in the Ophiuchus star-forming region. This sample is diverse in terms of SED Class and dust substructures \citep{Cieza2021}, and we found that this disc sample is also diverse in gas distributions. All discs show bright $^{12}$CO(2-1) emission, irrespective of their pre-main-sequence class, save for the Class II source ISO-Oph\,196, which is devoid of extended line emission except for a compact signal located inside its dust cavity. In this system, we detect compact emission in two channels only. A possible explanation for this might be that the disc dispersal process already started in this system, but these data must be interpreted with caution and further research is required.\\
Our channel maps show that most of the discs exhibit typical signatures of Keplerian rotation, except WLY\,2-63, which may be on the route of an outflow. We note that several sources display strong cloud contamination. Consequently, in several cases, the emission in channel maps differs from that expected from Keplerian rotation.\\
Finally, several disc surveys in different star-forming regions have reported that $^{12}$CO(2-1) emission tends to be more extended than the mm continuum emission \citep[e.g.][]{Ansdell2018, Villenave2021, Kurtovic2021, Rota2022}. This trend can be explained by two processes, a difference in optical depth, where the line optical depth is higher than in the dust, and grain growth and the posterior inward radial drift of millimeter-sized grains; besides, a size ratio $R_{gas}/R_{dust}>4$ can be an indicator of radial drift in discs \citep[][]{Hughes2008, Trapman2019, Facchini2019, Toci2021}. In this work, we measured the gas radius for the two sources less affected by emission from the cloud: RXJ1633.9-2442 and DoAr\,44, and we note that our measurements may be underestimated; in these cases, the radii are $\sim$\,3.0 and 1.4 times larger than the mm continuum respectively. These results are in agreement with previous observations; however, we notice that in other cases the gas size is clearly smaller than its mm continuum counterpart (e.g. ISO-Oph 54 and ISO-Oph 196). Most likely, this is due to the limitations of our observations in terms of MRS and sensitivity. Those sources are considerably fainter in $^{12}$CO than the rest of the sample.\\
Further and deeper spatially-resolved observations of molecular lines are needed to better characterize the gas component in the ODISEA discs. Moreover, new observations of optically thin lines of CO isotopologues, such as $^{13}$CO and C$^{18}$O are crucial to constrain the gas masses and surface density profiles. These observations would allow us to characterize the possible outflow in WLY\,2-63 \citep[e.g.][]{Ruiz-Rodriguez2017a, Ruiz-Rodriguez2017b}.

\subsubsection{Observed dust substructures in the ODISEA discs}
Dust rings and gaps are the most common substructures identified in protoplanetary discs and their origin is still unclear \citep[e.g.][]{vanderMarel2017, Andrews2020}. Different mechanisms are discussed in the literature regarding the origin of dust cavities, such as photoevaporative winds, planet-disc interactions, and dead zones \citep[e.g.][]{Pinilla2012, Flock2015, owen_2016, vanderMarel2017, Garate2021}. In this context, spatially resolved images of the gas are crucial to distinguish between different scenarios. Models of photoevaporation predict that both gas and dust are cleared simultaneously \citep[e.g.][]{Alexander2007}, and planet-disc interaction predicts an increase in gas density at the edge of the gap \citep[e.g.][]{Pinilla2012}. Besides, dead zones and magnetohydrodynamic winds can induce typical structures of transition discs, which are similar to those produced by embedded planets \citep[][]{Pinilla2016}.
In recent years, ALMA observations have revealed the presence of gas inside dust cavities for several discs \citep[e.g.][]{Casassus2013, Bruderer2014, Perez2015b, Canovas2015, VanDerMarel2015, VanDerMarel2016}. In our ODISEA sample, eight discs show evidence of gas inside dust cavities, therefore, the presence of planets may be responsible for the appearance of the observed gaps in those sources \citep[e.g.][]{VanDerMarel2016}. Moreover, the fact that the ODISEA long-baseline sample is strongly biased towards massive stars in Ophiuchus \citep[][]{Cieza2021} matches with the theory of planet-disc interaction because massive stars are considered good candidates to form planets \citep[][]{Pascucci2016}. At the same time, in the cases where the gas is present inside the dust cavities, the possibility of the photoevaporation mechanism is weakened. These results support evidence from previous studies of the SR\,24S \citep[][]{VanDerMarel2015,Pinilla_2017} and DoAr\,44 \citep[][]{VanDerMarel2016} systems. \citet{Pinilla2019} compared multiwavelength observations of SR\,24S with predictions of internal photoevaporation, dead zone, and planet-disc interaction models. The study concluded that internal photoevaporation and dead zone are inconsistent with the observations and that an embedded planet(s) is the most convincing mechanism for the formation of the SR\,24S' dust cavity.\\
New data of the $^{12}$CO(2$-$1) line, more sensitive and with higher spectral resolution would allow us to perform a more detailed analysis of the kinematics of the gas and, particularly, to explore the hypothesis of planet-disc interaction as an explanation for the formation of dust cavities in several ODISEA discs.

\subsubsection{Evidence of warped morphologies}
It is now well established from a variety of studies, that protoplanetary discs can develop warps \citep[e.g.][]{Marino2015, Benisty2017, Pinilla2018, Benisty2018, Sakai2019, Kraus2020}. Recent evidence suggests that inner disc misalignments are common in protoplanetary discs \citep{Ansdell2020}, and potential mechanisms causing warped morphologies include the interaction with an inclined and massive companion \citep[e.g.][]{Price2018, Zhu2019}. Recently, \citet{Young2021} found variations in the abundances of species in misaligned discs that could allow the identification of warped disc structures.
Interestingly, DoAr\,44, displays a twisted velocity structure in the moment 1 map (Fig.\,\ref{fig:doarPA}), which is a typical signature of misaligned or broken discs \citep[e.g.][]{Casassus2015, Facchini2018, Mayama2018}. This observational proof is consistent with the predictions made by \citet{Casassus2018}. We notice that the velocity field traced in Fig.\,\ref{fig:doarPA} is consistent with the prediction for the $^{12}$CO(6$-$5) line in \citet{Casassus2018} (see Fig. 8). Apart from a misaligned inner disc, another possible explanation for the twisted kinematic pattern in Fig.\,\ref{fig:doarPA} is the infall or accretion of material in the innermost part of the disc \citep[e.g.][]{Rosenfeld2014, Casassus2015, Mayama2018}. In addition, DoAr\,44 displays a difference in orientation between gas and dust, and it can thus be suggested that this disc is warped. Our findings strongly support the scenario of the misaligned inner disc for DoAr\,44. Therefore, further research should be undertaken to investigate its possible origin.

\section{Conclusions}
\label{sec:Conclusions}
We presented ALMA long-baseline observations of ten discs from the ODISEA survey and presented the first line observations for seven systems, WSB 82, ISO-Oph 17, WLY\,2-63, ISO-Oph 54, RXJ1633.9-2442, ISO-Oph 37, and ISO-Oph 196. We measured the gas disc size in two cases, also, in five cases computed the position angle, and compared the orientation of the gas traced by the $^{12}$CO(2-1) line with that from the continuum; The main results of this work can be summarized as follows:
\begin{enumerate}
\item Eight discs show evidence of gas inside inner dust cavities or gaps.
\item We found a significant difference between the orientation of the high-velocity $^{12}$CO gas, near the star, and the orientation of the dust in the outer ring for DoAr\,44.
\item A twisted kinematic structure is found in the central part of the moment 1 map of the $^{12}$CO(2-1) emission for DoAr\,44, which is an indicator of possible misalignment between inner and outer discs.
\item The channel maps of the flat spectrum source WLY\,2-63 provide evidence for a possible outflow, so further work is needed to figure out the origin of the observed features.
\item The $^{12}$CO(2-1) emission of ISO-Oph 196 displays a compact and non-symmetric emission located inside its cavity, which motivates future new observations of the components of this disc.
\end{enumerate}
These results motivate further and more sensitive observations in CO rotational lines and isotopologues of selected sources from the ODISEA survey.

\section*{Acknowledgements}
We thank the referee for their comments and suggestions, which improved the manuscript. This work was funded by ANID --Millennium Science Initiative Program-- Center Code NCN2021\_080. J.A., S.C. and L.C acknowledge support from Agencia Nacional de Investigaci\'on y Desarrollo de Chile (ANID) given by FONDECYT Regular grants 1211496 and 1211656.\\
This paper makes use of the following ALMA data: ADS/JAO.ALMA\#{\tt 2018.1.00028.S}. ALMA is a partnership of ESO (representing its member states), NSF (USA) and NINS (Japan), together with NRC (Canada), MOST and ASIAA (Taiwan), and KASI (Republic of Korea), in cooperation with the Republic of Chile. The Joint ALMA Observatory is operated by ESO, AUI/NRAO and NAOJ.

\section*{Data Availability}

The reduced ALMA data presented in this article are available upon
reasonable request to the corresponding author.


\bibliographystyle{mnras}
\bibliography{bibliography} 





\bsp	
\label{lastpage}
\end{document}